\documentclass[12pt]{article}
\usepackage{graphicx,psfrag,epsf}
\usepackage{caption}
\usepackage{hyperref}
\usepackage{enumerate}
\usepackage{natbib}
\usepackage{url} 
\usepackage{hhline,array,amsmath,graphicx}
\usepackage{pifont,latexsym,ifthen,rotating,calc}
\usepackage{siunitx}
\usepackage{etoolbox}
\usepackage{mathtools}
\usepackage{amsmath}
\usepackage{amssymb}
\usepackage{amsthm}
\usepackage{bbm}
\usepackage{pdflscape}
\usepackage{rotating}
\usepackage{bm}
\usepackage{enumitem}
\usepackage[mathscr]{euscript}
\usepackage{multirow}
\usepackage{xcolor}
\newcommand{\widesim}[2][1.5]{
  \mathrel{\overset{#2}{\scalebox{#1}[1]{$\sim$}}}
}

\newcommand{\blind}{0}

\begin{document}

\def\spacingset#1{\renewcommand{\baselinestretch}%
{#1}\small\normalsize} \spacingset{1}


\if0\blind
{
  \title{\bf Fiducial Confidence Intervals for Agreement Measures Among Raters Under a Generalized Linear Mixed Effects Model}
  \author{Soumya Sahu\\
    Department of Epidemiology and Biostatistics, University of Illinois Chicago\\
    Thomas Mathew\\
    Department of Mathematics and Statistics, University of Maryland Baltimore County
    and \\
    Dulal K. Bhaumik\\
    Department of Epidemiology and Biostatistics,\\
    Department of Psychiatry,
    University of Illinois Chicago}
  \maketitle
} \fi

\if1\blind
{
  \bigskip
  \bigskip
  \bigskip
  \begin{center}
    {\Large \bf Fiducial Confidence Intervals for Agreement Measures Among Raters Under a Generalized Linear Mixed Effects Model}
\end{center}
  \medskip
} \fi

\bigskip

\begin{abstract}
A generalization of the classical concordance correlation coefficient (CCC) is considered under a three-level design where multiple raters rate every subject over time, and each rater is rating every subject multiple times at each measuring time point. The ratings can be discrete or continuous. A methodology is developed for the interval estimation of the CCC  based on a suitable linearization of the model along with an adaptation of the fiducial inference approach. The resulting confidence intervals have satisfactory coverage probabilities and shorter expected widths compared to the interval based on Fisher’s Z-transformation, even under moderate sample sizes. Two real applications available in the literature  are discussed. The first application is based on a clinical trial to determine if various treatments are more effective than a placebo for treating knee pain associated with osteoarthritis. The CCC was used to assess agreement among the manual measurements of  the joint space widths on plain radiographs by two raters, and the computer-generated measurements of digitalized radiographs.  The second example is on a corticospinal tractography and the CCC was once again applied in order to evaluate the agreement between a well-trained technologist and a neuroradiologist regarding the measurements of fiber number in both the right and left corticospinal tracts. Other relevant
applications of our general approach are highlighted in  many areas including artificial intelligence.
\end{abstract}

\noindent%
{\it Keywords:}  Fiducial quantity, Fisher's Z-transformation, Hierarchical designs, Longitudinal data, Poisson regression. 
\vfill

\addtolength{\textheight}{.2in}%
\newpage
\spacingset{1.9} 
\section{Introduction}
\label{sec:intro}
\noindent Measures of agreement among different data-generating sources (referred to as raters), are needed to assess the acceptability of a new or generic process. A rater can be a chemist, a psychologist, a technique, or even a formula. Additionally, nowadays, in many instances computer
measurements have replaced human measurement, for example, image-based measures in medical research.  The books by \cite{linbook} and \cite{choudharybook} give book-length treatment of the criteria and methodologies for measuring agreement under various continuous and discrete scenarios.

The concordance correlation coefficient (CCC) due to \cite{lin1989} is perhaps the most popular and widely used criterion for assessing  agreement between two raters for measurements on a continuous scale.   
Over the years, the CCC has been extended and adopted for different scenarios, motivated by specific applications. Indeed, extensions of the CCC have been necessary due to several practical considerations. The measurements can be in dichotomous, polychotomous, ordinal, count, or continuous scales. Furthermore, a subject can be observed repeatedly over time by each of several raters. Thus, it is necessary to use models that include random effects and over-dispersion, and also accommodate the longitudinal nature of the data. Some references that investigate agreement measures based on models that take into account the above features of the data include \cite{carrasco2005}, \cite{king2007}, \cite{carrasco2010}, \cite{tsai2015}, \cite{tsai2018}, \cite{tsai2022}, \cite{bhaumik2021}, \cite{shi2022}, \cite{tsai2023}. We also note that when there are replicated measurements on each subject from several raters, the CCC can be defined in order to assess intra, inter and total agreement; see \cite{barnhart2005}. 

Clearly, point and interval estimation of the CCC are problems of considerable practical interest. 
The present article focuses on the interval estimation problem under a scenario where multiple raters rate every subject over time, and each rater is rating every subject multiple times at each time point. Therefore the design for each rater is structured as replicated measurements (level 1) nested within measuring time points (level 2) and the time points nested within subjects (level 3). A full agreement among the raters is established when: (i) they agree completely on how the fixed effects are affecting the outcomes i.e., the fixed effects are exactly the same over all the rater-specific models, and (ii) the variances in the outcomes due to random effects are the same within a rater and between  raters, as captured by rater specific variances, and the between rater covariances of the random effects, respectively.  Taking into account all of such considerations, an appropriate hierarchical linear mixed effects model for such designs is proposed by \cite{shi2022} for continuous ratings. The main structure of our model is motivated by their work; additionally, we have extended it to GLMM to make it adequate for binary, count or continuous responses. 

A standard approach for computing a confidence interval for the CCC is to use the asymptotic normality of the random variable resulting from Fisher's Z-transformation; see \cite{lin2002}, \cite{tsai2022}. Indeed, several of the above cited articles follow such an approach, and also evaluate the accuracy of the resulting confidence intervals in terms of maintaining the coverage probability. However, we shall pursue an alternative approach based on the idea of a {\it fiducial quantity}. The fiducial approach has seen a revival during the past several years; see \cite{hannig2009} and \cite{hannig2016} for detailed reviews. Earlier, the fiducial approach was developed and successfully applied to several problems by \cite{weerahandi1993}, where a fiducial quantity was referred to as a {\it generalized pivotal quantity}. We shall develop an ``approximate fiducial approach" under the GLMM after linearizing the model and using the asymptotic normality of the maximum likelihood estimators. As we shall show, this approach will result in confidence intervals that are preferable to those based on the Fisher's Z-transformation in two respects: the fiducial confidence intervals that we develop maintain the coverage probabilities better under reasonable sample size scenarios, and the expected width of the intervals are smaller compared to those of the intervals computed using the Fisher's Z-transformation. Here we note that in a simple setup of a bivariate normal distribution involving only two raters, the fiducial approach is developed in \cite{bhaumik2021} for the interval estimation of the CCC.    

The rest of the paper is organized as follows. In the next subsection we present two applications that fit into our modeling framework where the problem  of interest is inference concerning the CCC.  In Section 2, we introduce the GLMM and the agreement measures for such a set up. Section 3 focuses on the fiducial approach  and the application  of the fiducial idea in order to derive approximate fiducial quantities  for the parameters in the GLMM, and also for the CCC. Section 4 contains results from some simulation studies in order to assess the performance of the proposed approach for the interval estimation of the CCC. The two applications are reconsidered in Section 5 and the paper concludes with a discussion in Section 6.
We want to conclude this introduction by noting that a hypothesis testing approach appears inappropriate in the context of assessing agreement since there is no value that is a gold standard for the agreement measures. In view of this, the focus of our work is exclusively on interval estimation.

\subsection{Two Examples}

In this subsection, we shall briefly discuss two practical applications taken from the literature; these applications will be discussed later in the paper. The applications also provide further motivation for the interval estimation of the CCC, since no gold standard is available for the CCC, as noted earlier. The first application is in a scenario where the measurements are continuous, and in the second application, the measurements are counts. It is important to note that the sample sizes in both of the applications are small (38 and 10, respectively), which may raise questions regarding the applicability of the popularly used large sample-based (e.g. Fisher-Z)  confidence intervals.  A comparison between the large sample-based method and the proposed method is provided later in Section 4.  Details  of these examples, and the corresponding  modeling are given in Section 5.

\subsubsection{Example 1: The Glucosamine Arthritis Intervention Trial (GAIT)}

The GAIT is an NIH-funded, double-blind, five-arm randomized clinical trial designed to determine whether for the treatment of knee pain associated with osteoarthritis (OA) of the knee,  glucosamine, chondroitin sulfate, and/or the combination of glucosamine and chondroitin sulfate are more effective than placebo, and whether the combination is more effective than glucosamine or chondroitin sulfate alone (\cite{shi2022}, and cross references therein). A substudy was conducted in order to further examine the effects of these treatments on the Joint Space Width (JSW) over time. Two investigators: a rheumatologist with an extensive experience
in clinical investigation and previous experience in radiographic interpretation of clinical trials, and a musculoskeletal
radiologist, reviewed plain hard-copy radiographs and measured the JSW with calipers. The radiographs were read by these physician investigators independently without knowledge of the patient's name, participating clinic, treatment, and date of radiographs. An additional non-technical
rater used the computer program Mdisplay  to measure JSW on digitized images. 

The JWS of each subject was measured three times, and each rater rated three radiographs independently.   In addition,  after a time gap, each rater was asked to rate the same radiograph again for every subject that he/she rated before. Consequently each rater rated every subject three times with a repetition at each time point.  Thus we have longitudinal data under a three-level design where replicates (level 1) are nested within radiographs or x-ray plates (level 2), and plates are nested within subjects (level 3) (\cite{shi2022}). The data were right-skewed and a GLMM under the gamma distribution was used to analyze the study data.  The problem of interest  is to assess whether the manual measurements of JSW on plain radiographs is equally as reproducible as computer-generated measurement of the digitalized radiograph. Ideally,  between rater agreement should have been compared with the true value, known as the gold standard; however, in this example of measuring JSW - as is often the case, the gold standard is unknown.

\subsubsection{Reproducibility of Corticospinal Diffusion Tensor Tractography }

This application is on the reproducibility of corticospinal diffusion tensor tractography (DTT) in healthy subjects, and the goal of the study was the development of a guideline for comparison with the DTT in stroke patients before the longitudinal monitoring of the effects of stem cell therapy (\cite{tsai2018, tsai2022, tsai2023}). In this study, 10 healthy subjects were enrolled, and for each subject, the measurements of fiber number in the right and left corticospinal tracts derived from the diffusion tensor data were obtained by one well-trained technologist and a neuroradiologist. In addition, two scans in one session and
a third scan one week later were collected. Further, the bilateral corticospinal tracts in each
scan were reconstructed twice. A three-level longitudinal mixed-effects model for count data was used to analyze the study data, where two reconstructed scans each (level 1) were nested within the scans per session (level 2), and three scans were nested within subjects (level 3). The problem of interest was to compare the fiber  counts from the diffusion tensor data obtained by the technologist and neuroradiologist in order to evaluate the strength of the agreement. 

\section{A GLM Model and Measures of Agreement}

We start with a description of the model, followed by expressions for measures of agreement that are appropriate under the model. We consider a three-level design where multiple raters rate every subject over time, and each rater is rating every subject multiple times at each measuring time point.  The main structure of our model is motivated by the work of \cite{shi2022}, where the authors have introduced an appropriate hierarchical linear mixed effects model for the ratings. We have extended the model to the GLMM in order to make it adequate for binary, count, or continuous responses. Longitudinal data sets can be flexibly modeled using time splines, and subject-specific trajectories can be tackled via subject-specific random effects of the time splines. Due to such advantages, we have made our model flexible so as to include any basis function of the time splines, such as polynomial spline, natural spline, basis spline etc.

Let $y_{ijkl}$ denote the rating for the $i$-th subject at the $k$th replication for the $j$th time point evaluated by the $l$th rater, $i = 1,...,N$, $k = 1,..., K$, $j = 1,...,T$ and $l = 1,...,L$. Let $\bm{y}_{ijl}$ be a $K\times 1$ column vector consisting of the $K$ ratings for the $i$th subject at the $j$th time point by the $l$th rater, and write $\bm{y}_{il} = (\bm{y}^t_{i1l}, \bm{y}^t_{i2l},..., \bm{y}^t_{iTl})^t$. The conditional mean of $\bm{y}_{il}$ can be written as
\begin{equation}\label{mean}
    \bm{\mu}_{il} = E(\bm{y}_{il} | \bm{\Theta}) = h\left( \bm{X}_{i}\bm{\beta}_l + \sum \limits_{s = 0}^S \bm{z}_s\alpha_{il}^s + \tilde{\bm{\gamma}}_{il}\otimes \bm{1}_K \right) = h(\bm{\eta}_{il}),
\end{equation}
where $\bm{\Theta}$ denotes the set of all random effects, $h$ is the vector valued inverse link function for the  GLMM,  $\bm{\eta}_{il} = \bm{X}_{i}\bm{\beta}_l + \sum \limits_{s = 0}^S \bm{z}_s\alpha_{il}^s + \tilde{\bm{\gamma}}_{il}\otimes \bm{1}_K$, and $\otimes$ is the Kronecker product. Furthermore, 
$\bm{\alpha}_i^s = \left (
    \alpha_{i1}^s,
    \alpha_{i2}^s,
    \cdots ,
    \alpha_{iL}^s
    \right )^t$ represents $L\times 1$  subject-level random effects for $s$ = 0, 1, 2, ...., $S$,
    $\tilde{\bm{\gamma}}_{il} = \left (
    \gamma_{i1l},
    \gamma_{i2l},
    \cdots ,
    \gamma_{iTl}
    \right )^t $ and
$\bm{\gamma}_{ij} = \left (
    \gamma_{ij1},
    \gamma_{ij2},
    \cdots ,
    \gamma_{ijL} \right )^t$ denotes the $L\times 1$ subject-time random interaction effects.   
Additionally, $\bm{X}_i$ is the $KT \times d$ design matrix associated with the $d\times 1$ fixed effects parameters $\bm{\beta}_l$, $l=1,...,L$, $(\bm{z}_1, \bm{z}_2,..., \bm{z}_S)$ is the $KT\times S$ basis matrix for a time spline evaluated at the time vector $[(1, 2,..., T)^t\otimes \bm{1}_K]^t$, where ${\bf 1}_r$ denotes an $r\times 1$ vector of ones. Furthermore,  $\bm{z}_0 = \bm{1}_{KT}$, used for the random subject-level intercept consisting of other subject-level variations which are not functions of time. Fixed effects in the model usually include demographic information of the basis of the time splines.  In addition to the structure in (\ref{mean}) for the mean, the conditional variance is assumed to be 
\begin{equation}\label{var}
{\rm var}(\bm{y}_{il} | \bm{\Theta}) = {\rm diag}(\{\phi \zeta(\mu_{ijkl}), j = 1,...,T; k = 1,...,K\}),
\end{equation}
where $\phi$ is referred to as the dispersion parameter and $\zeta(\mu_{ijkl})$ is the conditional variance function associated with $h$. The distributional assumptions on the random effects are:

 Assumption 1: Subject-level random effects, $\bm{\alpha}_i^s \underset{i}{\widesim[1.5]{i.i.d.}} N_L (\bm{0}_L, \bm{\Sigma}_\alpha^s)$, where $\sigma_{\alpha pq}^s = cov(\alpha_{ip}^s, \alpha_{iq}^s)$ denotes $(p, q)$th element of $\bm{\Sigma}_\alpha^s$, $p = 1,...,L$, $q = 1,...,L.$ We also assume that the $\bm{\alpha}_i^s$'s are independent for $s = 0, 1,..., S$; i.e. $cov(\bm{\alpha}_i^s, \bm{\alpha}_i^{s'}) = \bm{0}.$ Thus, $\bm{\alpha}_i^s$'s are independent over $i$ as well as over $s$.
    
 Assumption 2: Subject-time random interaction effect, $\bm{\gamma}_{ij} \underset{i,j}{\widesim[1.5]{i.i.d.}} N_L (\bm{0}_L, \bm{\Sigma}_\gamma)$, where $\sigma_{\gamma pq} = {\rm cov}(\gamma_{ijp}, \gamma_{ijq})$ denotes $(p, q)$th element of  $\bm{\Sigma}_\gamma$, $p = 1,...,L$, $q = 1,...,L.$ 
    
 Assumption 3: Subject-level random effects and subject-time random interaction effects are independent, i.e. ${\rm cov}(\alpha_{il}^s, \gamma_{ijl}) = 0$, for $i = 1,...,n$, $j = 1,...,T$, $s = 1,...,S$. 
\subsection{Measures of Agreement}
\noindent For two raters with ratings $(Y_1, Y_2)$, \cite{lin1989} proposed the concordance correlation coefficient (CCC) as a measure of agreement, defined as $CCC_{\rm Lin} = 1 - \frac{E(Y_1-Y_2)^2}{E_I(Y_1-Y_2)^2}$, where $E_I()$ denotes expectation under the independence assumption, and $Y_1$ and $Y_2$ are both univariate. In the two-rater scenario, when observations are longitudinal, we should compare the pairs of ratings obtained by rating a subject at the same time point and corresponding to the same replication. Thus, let $\bm{y}_{1}$ and $ \bm{y}_{2}$ be longitudinal observations obtained by two raters, similar to $\bm{y}_{il}$ defined earlier. Then a $CCC$ measure for two raters can be defined as
$$CCC_2 = 1 - \frac{E[(\bm{y}_{1} - \bm{y}_{2})^t(\bm{y}_{1} - \bm{y}_{2})]}{E_I[(\bm{y}_{1} - \bm{y}_{2})^t(\bm{y}_{1} - \bm{y}_{2})]},$$
which is the same as the measure proposed by \cite{king2007}. Since $E_I[(\bm{y}_{1} - \bm{y}_{2})^t(\bm{y}_{1} - \bm{y}_{2})] = {\rm tr(var}(\bm{y}_{1})) + {\rm tr(var}(\bm{y}_{2})) - 2{\rm tr(cov}(\bm{y}_{1}, \bm{y}_{2})) + (E(\bm{y}_{1}) - E(\bm{y}_{2}))^t(E(\bm{y}_{1}) - E(\bm{y}_{2}))$, where ${\rm cov}(\bm{y}_{1}, \bm{y}_{2}) = 0$ under independence, we get
$$
CCC_2 = \frac{2{\rm tr(cov}(\bm{y}_{1}, \bm{y}_{2}))}{{\rm tr(var}(\bm{y}_{1})) + {\rm tr(var}(\bm{y}_{2})) + (E(\bm{y}_{1}) - E(\bm{y}_{2}))^t(E(\bm{y}_{1}) - E(\bm{y}_{2}))}.
$$ 

\noindent The above measure of agreement can be generalized for $L$ raters using the idea in \cite{tsai2018}, namely, $CCC_L = 1 - \frac{\sum \limits_{l = 1}^{L-1}\sum \limits_{l^\prime = l+1}^{L} E[(\bm{y}_{l} - \bm{y}_{l^\prime})^t(\bm{y}_{l} - \bm{y}_{l^\prime})]}{\sum \limits_{l = 1}^{L-1}\sum \limits_{l^\prime = l+1}^{L}E_L[(\bm{y}_{l} - \bm{y}_{l^\prime})^t(\bm{y}_{l} - \bm{y}_{l^\prime})]}$, where $\bm{y}_{l}$ denotes the longitudinal observations consisting of the ratings by the $l$th rater, $l = 1, 2, \cdots, L$. Similar to $CCC_2$ given above, a simplified expression can be derived as
\begin{equation}\label{CCCL}
\fontsize{11pt}{11pt}\selectfont
    CCC_L = \frac{2\sum \limits_{l=1}^{L-1} \sum \limits_{l'=l+1}^{L}{\rm tr(cov}(\bm{y}_{l}, \bm{y}_{l'}))}{(L-1)\sum \limits_{l=1}^L \sum \limits_{i=1}^N{\rm tr(var}(\bm{y}_{il})) + \sum \limits_{l=1}^{L-1} \sum \limits_{l'=l+1}^{L}(E(\bm{y}_{l}) - E(\bm{y}_{l'}))^t(E(\bm{y}_{l}) - E(\bm{y}_{l'}))}.
\end{equation}

\noindent In general, analytical evaluation of the expectation, variance, and covariance terms in the expression of $CCC_L$ can be challenging, and very often, they do not have closed-form expressions. We note that, under the models specified in (\ref{mean}) and (\ref{var}), ${\rm var}(y_{ijkl}) = {\rm var}(E(y_{ijkl}|\bm{\alpha}, \bm{\delta})) + E({\rm var}(y_{ijkl}|\bm{\alpha}, \bm{\delta})) = {\rm var}(\mu_{ijkl}) + \phi E(\zeta(\mu_{ijkl})$ and similarly, ${\rm cov}(y_{ijkl}, y_{ijkl'}|\bm{\alpha}, \bm{\delta})$ = ${\rm cov}((E(y_{ijkl}), E(y_{ijkl'}))|\bm{\alpha}, \bm{\delta}))) + E({\rm cov}(y_{ijkl}, y_{ijkl'})|\bm{\alpha}, \bm{\delta})$ = $  {\rm cov}(\mu_{ijkl}, \mu_{ijkl'})$,  where we have also used $E({\rm cov}(y_{ijkl}, y_{ijkl'})|\bm{\alpha}, \bm{\delta})=0$, in view of Assumptions 1 and  2. Thus, a simplified version of (\ref{CCCL}) can be obtained as 
\begin{equation}\label{CCCL-S}
\fontsize{11pt}{11pt}\selectfont
    CCC_L = \frac{2\sum \limits_{l=1}^{L-1} \sum \limits_{l'=l+1}^{L}\sum \limits_{i,j,k} {\rm cov}(\mu_{ijkl}, \mu_{ijkl'})}{(L-1)\sum \limits_{l=1}^L \sum \limits_{i,j,k} \{{\rm var}(\mu_{ijkl}) + \phi E(\zeta(\mu_{ijkl}))\} + \sum \limits_{l=1}^{L-1} \sum \limits_{l'=l+1}^{L} \sum \limits_{i,j,k} (E(\mu_{ijkl}) - E(\mu_{ijkl'}))^2}.
\end{equation}

\noindent One can use numerical integration or Monte Carlo integration to numerically compute the means, variances, and covariances in (\ref{CCCL-S}). The accuracy of such approximations for computing the $CCC_L$ is evaluated in Section 4 via simulations.

Based on the type of GLMM defined on $\bm{y}_{il}$, the expression in (\ref{CCCL-S}) can be further simplified. For example, for a linear mixed effects model, the link function $h(.)$ is the identity function, the dispersion parameter $\phi$ is usually denoted by  $\sigma^2$,  and the variance function $\zeta(.) = 1$, so that the error variance is $\sigma^2$ and the expression of the $CCC_L$ in (\ref{CCCL-S}) is
\begin{equation}\label{CCCL-LM}
\fontsize{9pt}{9pt}\selectfont
    CCC_L^{LM} = \frac{2\sum \limits_{l = 1}^{L-1}\sum \limits_{l^\prime = l+1}^{L} \left[\sum \limits_{s=0}^{S} \sigma_{\alpha l l^\prime}^s \bm{z_s}^t\bm{z_s} + KT\sigma_{\gamma l l^\prime} \right]}{(L-1)\sum \limits_{l=1}^L \left[\sum \limits_{s=0}^{S} \sigma_{\alpha ll}^s \bm{z_s}^t\bm{z_s} + KT\sigma_{\gamma ll} + KT \sigma^2\right] + \sum \limits_{l = 1}^{L-1}\sum \limits_{l^\prime = l+1}^{L} \left[  \sum \limits_{i=1}^N (\bm{\beta}_l - \bm{\beta}_{l^\prime})^t \bm{X}_i^t \bm{X}_i (\bm{\beta}_l - \bm{\beta}_{l^\prime}) \right]}.
\end{equation}

In the rest of this section, we shall focus on bounds for the $CCC_L$ in 
(\ref{CCCL-S}),  and conditions for attaining the bounds. Lower and upper bounds are given in the following theorem.

\medskip

\noindent{\it Theorem 1}. The agreement measure $CCC_L$ given in (\ref{CCCL-S}), under the model specified in (\ref{mean}) and (\ref{var}), satisfies the following bounds: 
\begin{equation}\label{bounds}
    -\left(1 + \frac{\sum \limits_{l=1}^L \sum \limits_{i,j,k}\phi E(\zeta(\mu_{ijkl}))}{\sum \limits_{l=1}^{L}\sum \limits_{i,j,k} {\rm var}(\mu_{ijkl})}\right)^{-1} \leq CCC_L \leq \left(1 + \frac{\sum \limits_{l=1}^L \sum \limits_{i,j,k}\phi E(\zeta(\mu_{ijkl}))}{\sum \limits_{l=1}^{L}\sum \limits_{i,j,k} {\rm var}(\mu_{ijkl})}\right)^{-1}.
\end{equation}

\medskip

\noindent{\it Proof:} 

\smallskip

Using the notations of the model given in (\ref{mean}) and (\ref{var}), we note that perfect agreement among the raters is obtained in the scenario where $\bm{\beta}_l = \bm{\beta}_{l^\prime}$, $\sigma_{\alpha l l^\prime}^s = \sigma_{\alpha l l}^s = \sigma_{\alpha l^\prime l^\prime}^s$ for all $s$ and $\sigma_{\gamma l l^\prime} = \sigma_{\gamma l l} = \sigma_{\gamma l^\prime l^\prime}$ for all $l \neq l^\prime$. In this case, the correlation between  $\alpha_{il}^s$ and $\alpha_{il'}^s$ is one (for all $l \neq l'$). The same is true for $\gamma_{ijl}$ and $\gamma_{ijl'}$. These results together with the assumption that the means of the random effects are zeros imply that the regression line between $\alpha^s_{il}$ and $\alpha^s_{il'}$ passes through the origin with an angle of $45^0$.  This implies that distribution-wise  $\alpha_{i1}^s$ and $\alpha_{i1'}^s$ are identically equal for all $i$ and $s$, and similarly $\gamma_{ijl}$ and $\gamma_{ijl'}$ are identically equal for all $i$ and $j$. Hence,  $E(\mu_{ijkl}) = E(\mu_{ijkl'})$ and ${\rm cov}(\mu_{ijkl}, \mu_{ijkl'}) = {\rm var}(\mu_{ijkl})$ for all $j, k, l, l'$ and the $CCC_L$ can be expressed as $\left(1 + \frac{\sum \limits_{l=1}^L \sum \limits_{i,j,k}\phi E(\zeta(\mu_{ijkl}))}{\sum \limits_{l=1}^{L}\sum \limits_{i,j,k} {\rm var}(\mu_{ijkl})}\right)^{-1}$,  which is an upper bound for the $CCC_L$ for given values of $\bm{\beta}_l$, $\sigma_{\alpha l l}^s$, $\sigma_{\gamma l l}$, for all $s$ and $l$. Similarly, when $\bm{\beta}_l = \bm{\beta}_{l^\prime}$, $-\sigma_{\alpha l l^\prime}^s = \sigma_{\alpha l l}^s = \sigma_{\alpha l^\prime l^\prime}^s$ for all $s$ and $-\sigma_{\gamma l l^\prime} = \sigma_{\gamma l l} = \sigma_{\gamma l^\prime l^\prime}$ for all $l \neq l^\prime$, a lower bound for the $CCC_L$  is $\left(-1 - \frac{\sum \limits_{l=1}^L \sum \limits_{i,j,k}\phi E(\zeta(\mu_{ijkl}))}{\sum \limits_{l=1}^{L}\sum \limits_{i,j,k} {\rm var}(\mu_{ijkl})}\right)^{-1}$ for given $\bm{\beta}_l$, $\sigma_{\alpha l l}^s$, $\sigma_{\gamma l l}$ for all $s$ and $l$. \qedsymbol{} 

\medskip

Since we do not have any general  expression with a closed-form for the $CCC_L$, we would like to  use the expression given in  (\ref{CCCL-LM}) for  $CCC_L^{LM}$ under a linear mixed-effects model in order to get a better understanding for the development of a criteria for attaining the bound  in terms of the model parameters. We can conclude the following:

\begin{itemize}
\item When the fixed parameter vectors are the same (i.e. $\bm{\beta}_l = \bm{\beta}_{l^\prime}(=\bm{\beta}$, say)),    and all variance components of random effects of  raters are also the same with their covariance  (i.e. $\sigma_{\alpha l l^\prime}^s = \sigma_{\alpha l l}^s = \sigma_{\alpha l^\prime l^\prime}^s (=\sigma_{\alpha}^s, say)$ for all $s$ and $\sigma_{\gamma l l^\prime} = \sigma_{\gamma l l} = \sigma_{\gamma l^\prime l^\prime}$ for all $l \neq l^\prime$)  and all the random variations are explained by the random effects present in the model (i.e. no unexplained or error variation, $\sigma^2 = 0$ for all $j, l$), the measure of agreement can achieve the upper limit $1$, i.e. $CCC_L^{LM} = 1$.  
\item If $\sigma_{\alpha l l^\prime}^s = 0$ for all $s$ and $\sigma_{\gamma l l^\prime} = 0$ for all $l \neq l^\prime$, $CCC_L^{LM} = 0.$ Assuming that all raters evaluate the study subjects independently, i.e., they don't have any association while evaluating the explained fixed and random variations in the measurements, the measure of association takes the value zero.
\item If $\bm{\beta}_l = \bm{\beta}_{l^\prime}(=\bm{\beta}, say)$, $-\sigma_{\alpha l l^\prime}^s = \sigma_{\alpha l l}^s = \sigma_{\alpha l^\prime l^\prime}^s (=\sigma_{\alpha}^s, say)$ for all $s$ and $-\sigma_{\gamma l l^\prime} = \sigma_{\gamma l l} = \sigma_{\gamma l^\prime l^\prime} (=\sigma_{\gamma}, say)$ for all $l \neq l^\prime$ and $\sigma^2 = 0$ for all $j, l$, $CCC_L^{LM} = -1$.    
\end{itemize}

The exact value of the $CCC_L$ for a given data set can fall short of the limits 1 or -1 even under perfect agreement or disagreement scenarios. The deviation of the $CCC_L$ from the limit -1 or 1 depends on the magnitude of $\frac{\sum \limits_{l=1}^L \sum \limits_{i,j,k}\phi E(\zeta(\mu_{ijkl}))}{\sum \limits_{l=1}^{L}\sum \limits_{i,j,k} {\rm var}(\mu_{ijkl})}$, and this in turn depends on the nature of the GLMM under consideration; in particular, it depends on the link function $h(.)$, dispersion parameter $\phi$ and the variance function $\zeta(.)$. In practice, it is suggested to compute the bounds in (\ref{bounds}) for the $CCC_L$ using the estimated parameters before performing the interval estimation so that the width of the confidence interval can be compared with the difference between the upper and lower bounds  of the $CCC_L$ given in (\ref{bounds}). 

\section{Interval Estimation of $CCC_L$}
\label{sec:verify}

A standard approach for the interval estimation of the CCC is to use asymptotic normality of the Fisher's Z-transformed version of the CCC; see, for example, \cite{tsai2018}.  However, this approach requires a large sample to get satisfactory performance in terms of providing coverage probabilities close to the target value. In a simple bivariate normal model, \cite{bhaumik2021}  used a {\it fiducial approach} for the interval estimation of the CCC. Here we shall pursue the approach of fiducial inference, which has found numerous applications, resulting in accurate inference even for some small sample problems; see \cite{hannig2009} and \cite{hannig2016}  for detailed reviews and a multitude of  applications. Fiducial inference for  CCC requires the construction of a fiducial quantity for the CCC, and percentiles of the fiducial quantity provide confidence limits. Fiducial quantities were earlier introduced (under the term generalized pivotal quantities) by \cite{weerahandi1993}. In our application, the fiducial quantities obtained will only be approximate fiducial quantities, constructed using a transformation of the GLMM. Later we shall compare the confidence intervals so obtained with those obtained using the Fisher's Z transformation.

\subsection{A Transformed Model}

\noindent We shall transform the responses in the GLMM so that the transformed responses can be assumed to follow an approximate LMM with the
same fixed and random effects, along with an additional error term. For this, let  $\hat{\bm{\beta}}_l$, $\hat{\alpha}_{il}^s$ and $\hat{\tilde{{\bm{\gamma}}}}_{il}$ denote consistent estimates of $\bm{\beta}_l$, $\alpha_{il}^s$ and $\tilde{{\bm{\gamma}}}_{il}$, respectively (\cite{Jiang1996}). Using a Taylor series expansion of the function $h$ in (\ref{mean}) with respect to both fixed and random effects, as done in \cite{dang2008} and \cite{amatya2018}), the response vector $\bm{y}_{il}$ can be approximated as follows: 
\begin{equation}\label{approx}
    \bm{y}_{il} = \hat{\bm{\mu}}_{il} + (\mathcal{G}^{\prime}(\hat{\bm{\mu}}_{il}))^{-1} \left[ \bm{X}_{i}(\bm{\beta}_l - \hat{\bm{\beta}}_l) + \sum \limits_{s = 0}^S \bm{z}_s(\alpha_{il}^s - \hat{\alpha}_{il}^s) + (\tilde{{\bm{\gamma}}}_{il} - \hat{\tilde{{\bm{\gamma}}}}_{il})\otimes \bm{1}_K \right ] + \bm{\epsilon}_{il},
\end{equation}
where $\mathcal{G}^{\prime}(\bm{\mu}_{il}) = [\frac{\delta h(\bm{\eta}_{il})}{\delta\bm{\eta}_{il}^t}]^{-1}$ is a diagonal matrix that is a function of $\bm{\mu}_{il}$ via $h$, and $\bm{\epsilon}_{il}$ is assumed to be a vector of independent errors, whose components  $\epsilon_{ijkl}$ have zero means and conditional variance ${\rm var}(\epsilon_{ijkl}) = \phi \zeta(\mu_{ijkl})$. 
It should be noted that such a linearization is already available in the literature, and also in widely used statistical software.  For example, SAS PROC GLIMMIX uses linearization methods to estimate model parameters of a GLMM, implementing penalized quasi-likelihood (PQL)  or marginal quasi-likelihood (MQL) methods ( \cite{dang2008}, \cite{Wolf1993}, \cite{SasG2006}). The  popular repackage `lme4'  employs the linear approximation  (\ref{approx}) in the penalized iteratively reweighted least squares (PIRLS) algorithm for  estimating  model parameters in a GLMM; see \cite{bates2007linear}  (page 29).

Let $\bm{y}_{il}^* = \hat{\bm{\eta}}_{il} + \mathcal{G}^{\prime}(\hat{\bm{\mu}}_{il})(\bm{y}_{il} - \hat{\bm{\mu}}_{il})$ and $\bm{\epsilon}_{il}^* = \mathcal{G}^{\prime}(\hat{\bm{\mu}}_{il})\bm{\epsilon}_{il}$, where $\bm{y}_{il}^*$ is a vector of ``pseudo-observations". Then, (\ref{approx}) can be re-expressed as
\begin{equation}\label{LMM}
    \bm{y}_{il}^* = \bm{X}_{i}\bm{\beta}_l + \sum \limits_{s = 0}^S \bm{z}_s\alpha_{il}^s + \tilde{\bm{\gamma}}_{il}\otimes \bm{1}_K + \bm{\epsilon}_{il}^*,
\end{equation}
which can be viewed as an LMM for the continuous pseudo-observation vectors $\bm{y}_{il}^*$ and random error vectors $\bm{\epsilon}_{il}^*$, both having dimensions $KT\times 1$, with the assumption $\bm{\epsilon}_{il}^* \underset{i,l}{\widesim[1.5]{i.i.d.}} N (\bm{0}, \phi \mathcal{G}^{\prime}(\bm{\mu}_{il}) \zeta(\bm{\mu}_{il}) \mathcal{G}^{\prime}(\bm{\mu}_{il}))$. Note that, $\zeta(\bm{\mu}_{il})$ is a $KT \times KT$ diagonal matrix with entries $\zeta(\mu_{ijkl})$. The conditional variance of $\bm{\epsilon}_{il}^*$ is computed using the continuity property of $\mathcal{G}^{\prime}(\bm{\mu}_{il})$ and consistency property of $\hat{\bm{\mu}}_{il}$.

We shall now explore the construction of fiducial quantities for the parameters in the GLMM. For this, we are going to use a construction given in \cite{krishnamoorthy2007} of a fiducial quantity for the entire covariance matrix, say $\bm{\Sigma}$, of a multivariate normal distribution; the construction is reproduced in Section A.1 of the Appendix attached to the supplementary material. We observe that the fiducial quantity, say  $\widetilde{\bm{\Sigma}}$, is a function of the observed data and pivot statistics such that (i) given the observed data, the distribution of $\widetilde{\bm{\Sigma}}$ is free of any unknown parameters, and (ii) if the random variables in the pivot statistics are replaced by the corresponding observed data, $\widetilde{\bm \Sigma}$ simplifies to $\bm\Sigma$. 
We want to point out that this is not the most general definition of a fiducial quantity, and we refer to \cite{hannig2009} for a general treatment.  However, this definition will serve our purpose. 

The parameters of interest in the GLMM specified in (\ref{mean}) and (\ref{var}) are the fixed effects $\bm{\beta}_l$, for $l=1,...,L$, covariance matrices for subject-level random effects, namely, $\bm{\Sigma}_\alpha^s$, for $s = 1,...,S$, and the covariance matrix for the subject-time random interaction effect, namely, $\bm{\Sigma}_\gamma$. Additionally,  we  have the dispersion parameter $\phi$.
We shall use REML estimates; the asymptotic properties of such estimates are discussed in \cite{Jiang1996}. However, other estimates satisfying similar asymptotic properties can also be used. Further, we shall take advantage of the properties of the LMM defined in (\ref{LMM}) which is a transformed version of the GLMM in (\ref{mean}). Hereafter, we shall refer to $\bm{y}_{il}^*$ as observed data that follows the LMM in (\ref{LMM}).  Estimates of the model parameters are obtained using the model in (\ref{mean}) only, and these are then used to obtain the LMM in (\ref{LMM}); no point estimation is performed using (\ref{LMM}). In particular, we have used the consistency and asymptotic normality of the estimates from model (\ref{mean}). We divide the discussion that follows into two parts: the first part focuses on the construction of fiducial quantities for the covariance matrices of random effects. These are then used in the second part in order to construct fiducial quantities for the fixed effects.
The fiducial quantity for $\phi$ will be denoted by $\widetilde{\phi}$. For the time being, we will assume that $\widetilde{\phi}$ is available; its construction will be addressed later.

Before we give the technical derivation of the fiducial quantities for the parameters in the GLMM, we shall give a brief explanation of the approach we shall pursue, using the approximate LMM (\ref{LMM}). We shall first consider the conditional multivariate normal distribution of $((\bm{\alpha}_i^s)^t, \sum \limits_{j=1}^T\bm{\gamma}_{ij}^t)^t$, conditionally given all of the $\bm{y}_{il}^*$. Let $(\bm{\mu}_{\alpha_s i}^t, {\bm{\mu}}_{\gamma i}^t)^t$ denote the corresponding conditional mean vector, which is a function of the GLMM parameters. Let $(\hat{\bm{\mu}}_{\alpha_s i}^t, \hat{\bm{\mu}}_{\gamma i}^t)^t$ denote the estimated conditional mean after replacing the unknown parameters with consistent estimates. We now appeal to the asymptotic normality of   $(\hat{\bm{\mu}}_{\alpha_s i}^t, \hat{\bm{\mu}}_{\gamma i}^t)^t$.  We will actually consider $(\hat{\bm{\mu}}_{\alpha_0 i}, \hat{\bm{\mu}}_{\alpha_1 i}, ..., \hat{\bm{\mu}}_{\alpha_S i}, \hat{\bm{\mu}}_{\gamma i})$  for $i=1,...,N$ as $N$ i.i.d. observations from a multivariate normal distribution having a zero mean vector and an arbitrary covariance matrix, say $\Delta$. The construction given in \cite{krishnamoorthy2007} can be used to obtain a fiducial quantity, say $\widetilde{\Delta}$,  for this arbitrary covariance matrix. However, the asymptotic covariance matrix of $(\hat{\bm{\mu}}_{\alpha_0 i}, \hat{\bm{\mu}}_{\alpha_1 i}, ..., \hat{\bm{\mu}}_{\alpha_S i}, \hat{\bm{\mu}}_{\gamma i})$ is a known function of the parameters in the GLMM. Hence the elements of  $\widetilde{\Delta}$ can be thought of as known functions of the fiducial quantities  of the parameters in the GLMM. The latter can now be obtained by doing least squares since the number of parameters in $\Delta$ is more than the number of parameters in the GLMM. The next subsection gives the details of the idea just outlined. We want to  point out that the observed data used in the derivation of the fiducial quantities are $(\hat{\bm{\mu}}_{\alpha_0 i}, \hat{\bm{\mu}}_{\alpha_1 i}, ..., \hat{\bm{\mu}}_{\alpha_S i}, \hat{\bm{\mu}}_{\gamma i})$, $i=1,...,N$. 
We also want to emphasize that the fiducial quantities that we derive are all approximate fiducial quantities, since they rely on the approximate LMM (\ref{LMM}), and on the consistency and asymptotic normality of the parameter estimates of the GLMM. However, we shall simply refer to them as fiducial quantities.

\subsubsection{Fiducial Pivot Statistics for the Covariance Matrices of  Random Effects}

The expression of $CCC_L$ given in (\ref{LMM}), in terms of variance components and fixed parameters, is explicit for the linear mixed effects model.  However, for the GLMM when  $CCC_L$ is expressed using pseudo outcomes (i.e. $Y^*$), the denominator in (\ref{LMM}) is a complex function of the variance components and the fixed effects parameters, with no explicit expression.  In what follows, we develop  fiducial quantities for the unknown fixed-effects parameters, and for the multivariate variance components of the random effects, leading to a fiducial quantity  for the $CCC_L$.  Denote, $\bm{Y}_i^* = \begin{pmatrix}
{\bm{y}_{i1}^*}^t & \hdots & {\bm{y}_{iL}^*}^t
\end{pmatrix}^t$, $\bm{\mathcal{X}}_i = \bm{I}_L \otimes \bm{X}_i$, $\bm{\beta} = \begin{pmatrix}
{\bm{\beta}_{1}}^t & \hdots & {\bm{\beta}_{L}}^t
\end{pmatrix}^t$ and $\bm{\Sigma}_{Y^*}$ = (${\Sigma_{Y^*}^{ll'}})$, $l,l' =1,2, \cdots, L$, where
 
  ${\Sigma_{Y^*}^{ll}} = {\rm var}(\bm{y}_{il}^*) = \sum \limits_{s=0}^S \sigma_{\alpha ll}^s\bm{z}_s\bm{z}_s^t + \sigma_{\gamma ll}\bm{I}_T \otimes \bm{J}_K + E_{\bm{\Theta}}[\phi \mathcal{G}^{\prime}(\bm{\mu}_{il}) \zeta(\bm{\mu}_{il}) \mathcal{G}^{\prime}(\bm{\mu}_{il})]$ and\\ ${\Sigma_{Y^*}^{ll^\prime}} = {\rm cov}(\bm{y}_{il}^*, \bm{y}_{il^\prime}^*) = \sum \limits_{s=0}^S \sigma_{\alpha ll^\prime}^s\bm{z}_s\bm{z}_s^t + \sigma_{\gamma ll^\prime}\bm{I}_T \otimes \bm{J}_K$ for $l\neq l^\prime$, $l=1,...,L$, $l^\prime=1,...,L$.

Using the properties of the LMM in (\ref{LMM}) and  Assumption 1 of (\ref{mean}), for every $s$ in $\{0,1,2,...,S\}$,
$$\begin{pmatrix}
\bm{\alpha}_i^s\\
\bm{Y}_i^*
\end{pmatrix} \sim N \left(\begin{pmatrix}
\bm{0}_L\\
\bm{\mathcal{X}}_i \bm{\beta}
\end{pmatrix}, \begin{pmatrix}
\bm{\Sigma}_\alpha^s & \bm{\sigma}^s_\alpha\\
{\bm{\sigma}^s_\alpha}^t & \bm{\Sigma}_{Y^*}
\end{pmatrix} \right),$$
where, $\bm{\sigma}^s_\alpha = {\rm cov}(\bm{\alpha}_i^s, \bm{Y}_i^*) = \begin{pmatrix}
\bm{\sigma}^s_{\alpha(1)} & \hdots & \bm{\sigma}^s_{\alpha(L)}
\end{pmatrix}$, 
$\bm{\sigma}^s_{\alpha(l)} = {\rm cov}(\alpha_{il}^s, \bm{Y}_i^*)$ is a column vector of dimension $KTL$, 
and typical elements of this vector are cov$(\alpha_{il}^s, y^*_{ikjl})=z_s\sigma^s_{\alpha ll}$  and cov$(\alpha_{il}^s, y^*_{ikjl'})=z_s\sigma^s_{\alpha ll'}$; both expressions remain the same for all  $k=1,2,..K$, $j=1,2,..T$, but vary over  $l=1,2,..L$. Thus, $\bm{\sigma}^s_{\alpha(l)}$ = $(\bm{1}^t_{KT}z_s\sigma^s_{\alpha l1}...\bm{1}^t_{KT}z_s\sigma^s_{\alpha ll}...\bm{1}^t_{KT}z_s\sigma^s_{\alpha lL})^t$.

Similarly, using the properties of the LMM in (\ref{LMM}) and Assumption 2 of (\ref{mean}),
$$\begin{pmatrix}
\sum \limits_{j=1}^T\bm{\gamma}_{ij}\\
\bm{Y}_i^*
\end{pmatrix} \sim N \left(\begin{pmatrix}
\bm{0}_L\\
\bm{\mathcal{X}}_i \bm{\beta}
\end{pmatrix}, \begin{pmatrix}
T\bm{\Sigma}_\gamma & \bm{\sigma}_\gamma\\
{\bm{\sigma}_\gamma}^t & \bm{\Sigma}_{Y^*}
\end{pmatrix} \right),$$
where $\bm{\sigma}_\gamma = \sum \limits_{j=1}^T \bm{\sigma}_{\gamma_j}$ and 
$\bm{\sigma}_{\gamma_j} = {\rm cov}(\bm{\gamma}_{ij}, \bm{Y}_i^*) = \begin{pmatrix}
\bm{\sigma}_{\gamma_j(1)} & \hdots & \bm{\sigma}_{\gamma_j(L)}
\end{pmatrix}$ and  each  $\bm{\sigma}_{\gamma_j(l)}$ is a column vector of dimension $KTL$.  Using  Assumptions 1-3, we compute the elements of the vector $\bm{\sigma}^t_{\gamma_j(l)}$ as cov$(\gamma_{ijl}, y^*_{ijkl'}) =\sigma_{\gamma ll'}$, and cov$(\gamma_{ijl}, y^*_{ij'kl'}) =0$ for all $j \neq j'$.  Thus in  the vector $\bm{\sigma}^t_{\gamma_j(l)}$, for each $l' = 1, 2, \cdots, L$, the element $\sigma_{\gamma ll'}$ is repeated $K$ times (when $j=j')$, and the element $0$ is repeated $K(T-1)$ times (when $j \neq j')$. Based on this observation we can write:
$\bm{\sigma}^t_{\gamma_j(l)}$ = $(a^t_{jl1} \; a^t_{jl2} \; ... \; a^t_{jlL})$ , where $a^t_{jll'}  =(\bm{0}^t_{K(j-1)} \; \; \bm{1}^t_K\sigma^t_{\gamma ll'} \; \; \bm{0}^t_{K(T-j)}).$ Thus we obtain $\bm{\sigma}_{\gamma_j(l)} = \begin{pmatrix}
\sigma_{\gamma l1} & \hdots & \sigma_{\gamma lL}
\end{pmatrix}^t \otimes \bm{e}_j^T \otimes \bm{1}_{K}$ and $\bm{\sigma}_\gamma = \sum \limits_{j=1}^T \bm{\sigma}_{\gamma_j} = \begin{pmatrix}
\sigma_{\gamma l1} & \hdots & \sigma_{\gamma lL}
\end{pmatrix}^t \otimes \bm{1}_{KT}$.
\\ 
\noindent Using the property of the conditional distribution of a  multivariate normal,
$\bm{\mu}_{\alpha_s i} = E[\bm{\alpha}_i^s | \bm{Y}_i^*] = \bm{\sigma}^s_\alpha \bm{\Sigma}_{Y^*}^{-1}(\bm{Y}_i^* - \bm{\mathcal{X}}_i\bm{\beta})$ for all $s=0,...,S$ and $\bm{\mu}_{\gamma i} = \sum \limits_{j=1}^TE[\bm{\gamma}_{ij} | \bm{Y}_i^*] = \bm{\sigma}_\gamma \bm{\Sigma}_{Y^*}^{-1}(\bm{Y}_i^* - \bm{\mathcal{X}}_i\bm{\beta})$. By plugging in the consistent estimators of fixed effects and variance components, estimates of $\bm{\mu}_{\alpha_s i}$ and $\bm{\mu}_{\gamma i}$ can be obtained, say, $\hat{\bm{\mu}}_{\alpha_s i}$ and $\hat{\bm{\mu}}_{\gamma i}$, respectively, for $i=1,...,N$, $j=1,...,T$. Using the consistency property of these estimates and Slutsky's theorem, we conclude that $(\hat{\bm{\mu}}_{\alpha_s i}^t, \hat{\bm{\mu}}_{\gamma i}^t)^t$ is asymptotically normal with a zero mean vector and variance-covariance parameters given by
\begin{eqnarray}\label{mu_alpha}
    &&{\rm var}(\hat{\bm{\mu}}_{\alpha_s i})= \bm{\sigma}^s_\alpha \bm{\Sigma}_{Y^*}^{-1} {(\bm{\sigma}^s_\alpha})^t,  \ {\rm var}(\hat{\bm{\mu}}_{\gamma i}) = \bm{\sigma}_\gamma \bm{\Sigma}_{Y^*}^{-1} ({\bm{\sigma}_\gamma})^t,
    \nonumber\\
&&{\rm cov}(\hat{\bm{\mu}}_{\alpha_s i}, \hat{\bm{\mu}}_{\alpha_{s'} i}) = \bm{\sigma}^s_\alpha \bm{\Sigma}_{Y^*}^{-1} ({\bm{\sigma}^{s'}_\alpha})^t, 
\ \mbox{cov}(\hat{\bm{\mu}}_{\alpha_s i}, \hat{\bm{\mu}}_{\gamma_i}) = \bm{\sigma}^s_\alpha \bm{\Sigma}_{Y^*}^{-1} ({\bm{\sigma}_\gamma})^t,  
\end{eqnarray}
for all $i$ and for $s, s'=0,...,S$. As noted earlier, we shall treat $(\hat{\bm{\mu}}_{\alpha_0 i}, \hat{\bm{\mu}}_{\alpha_1 i}, ..., \hat{\bm{\mu}}_{\alpha_S i}, \hat{\bm{\mu}}_{\gamma i})$  for $i=1,...,N$ as $N$ i.i.d. observations from the joint $(S + 2)L$-variate normal distribution and find the pivotal quantities for $\sigma_{\alpha ll'}^s$ for $s=0,...,S$ and $\sigma_{\gamma ll'}$ for $l, l' =1,...,L, l\leq l'$ in the following two steps:\\
{\it Step 1 \; Calculate pivots of the covariance matrix parameters of the normally distributed random variable $(\hat{\bm{\mu}}_{\alpha_0 i}, \hat{\bm{\mu}}_{\alpha_1 i}, ..., \hat{\bm{\mu}}_{\alpha_S i}, \hat{\bm{\mu}}_{\gamma i})$:} Utilizing (\ref{mu_alpha}) and the asymptotic joint normality of $(\hat{\bm{\mu}}_{\alpha_0 i}, \hat{\bm{\mu}}_{\alpha_1 i}, ..., \hat{\bm{\mu}}_{\alpha_S i}, \hat{\bm{\mu}}_{\gamma i})$  and  the method described in  Section A.1 in the Appendix available in the supplementary material, joint pivot statistics can be obtained  for the components of: (i) variance of $\hat{\bm{\mu}}_{\alpha_s i}$, i.e.,  $\bm{\sigma}^s_\alpha \bm{\Sigma}_{Y^*}^{-1} ({\bm{\sigma}^s_\alpha})^t$, for $s=0,...,S$; total of $(S+1)L(L+1)/2$ components, (ii) variance of $\hat{\bm{\mu}}_{\gamma i}$, i.e., $\bm{\sigma}_\gamma \bm{\Sigma}_{Y^*}^{-1} ({\bm{\sigma}_\gamma})^t$; a total of $L(L+1)/2$ components, (iii) covariance among $\hat{\bm{\mu}}_{\alpha_s i}$, i.e., $\bm{\sigma}^s_\alpha \bm{\Sigma}_{Y^*}^{-1} ({\bm{\sigma}^{s'}_\alpha})^t$, for $s,s'=0,...,S, s<s'$; a  total of $(S+1)(S+2)L^2/2$ components, and (iv) covariance between $\hat{\bm{\mu}}_{\alpha_s i}$ and $\hat{\bm{\mu}}_{\gamma i}$, i.e., $\bm{\sigma}^s_\alpha \bm{\Sigma}_{Y^*}^{-1} ({\bm{\sigma}_\gamma})^t$ for $s=0,...,S$; a total of $(S+1)L^2$ components.\\
{\it Step 2 \; Transform the pivots in Step 1 to the pivots of the variance parameters of the models in (\ref{mean}) and (\ref{var}):} The variance and covariance parameters of the models in  (\ref{mean}) and (\ref{var}), comprising of a total of $(S+2)L\left[(S+2)L+1\right]/2$ components, are functions of variance parameters $\sigma_{\alpha ll'}^s$ for $s=0,...,S$ and $\sigma_{\gamma ll\prime}$ for $l, l' =1,...,L, l\leq l'$, which are the elements of  $\bm{\Sigma}_\alpha^s$ for $s=0,...S$ and $\bm{\Sigma}_\gamma$ respectively under the model (\ref{mean}) and (\ref{var}). Ideally, we want to equate these $(S+2)L\left[(S+2)L+1\right]/2$ analytical expressions to their respective fiducial pivotal quantities found in Step 1 and solve for the $L(L+1)(S+2)/2$ variance and covariance parameters under the model (\ref{mean}) and (\ref{var}). Unfortunately, the number of equations is more than the number of parameters. To overcome this difficulty, a \textit{least-squares} approach is recommended. In this approach, for each of the $(S+2)L\left[(S+2)L+1\right]/2$ fiducial quantities from Step 1, we minimize the sum of the squared differences between each fiducial pivotal quantity and the corresponding parametric function. Thus, the problem comes down to an optimization problem which can be solved using an \textit{inexact Newton method} (see Chapter 7 in \cite{optimizationbook}). The optimization algorithm is robust for non-convexity and performs the Newton step in an \textit{inexact} way with a Hessian-free approach (computation of Hessian only needs the computational cost of computing the first derivative) which makes the algorithm very efficient even for solving large-scale problems. The algorithm performs line search with adaptive step size computation and effectively uses the second derivative information, calculated by conjugate gradient, to find the descent direction.

 Note that the fiducial quantities we have derived are expected to be a function of the dispersion parameter $\phi$. If so, $\phi$ must be replaced by a fiducial pivot statistic in the system of non-linear equations prior to solving them. Also, since the covariance matrices are nonnegative definite (nnd), the solution needs to be computed under the nnd constraint. In order to guarantee nonnegative definiteness, we have used the log-Cholesky decomposition (\cite{bates1996}) of the covariance matrices so that each transformed variance component can take any value on the real line. As this is a one-to-one transformation, the covariance matrices can be easily recovered after the solution is obtained.

When the covariances among $\hat{\bm{\mu}}_{\alpha_s i}$, i.e. the quantities $\bm{\sigma}^s_\alpha \bm{\Sigma}_{Y^*}^{-1} ({\bm{\sigma}^{s'}_\alpha})^t$ for $s,s'=0,...,S, s<s'$ and the covariance between $\hat{\bm{\mu}}_{\alpha_s i}$, $\hat{\bm{\mu}}_{\gamma i}$, i.e. the quantity  $\bm{\sigma}^s_\alpha \bm{\Sigma}_{Y^*}^{-1} ({\bm{\sigma}_\gamma})^t$ for $s=0,...,S$, are all very small compared to the respective variances, i.e. $\bm{\sigma}^s_\alpha \bm{\Sigma}_{Y^*}^{-1} ({\bm{\sigma}^{s}_\alpha})^t$ for $s=0,...,S$, and $\bm{\sigma}_\gamma \bm{\Sigma}_{Y^*}^{-1} ({\bm{\sigma}_\gamma})^t$, one can avoid using the joint distribution of $(\hat{\bm{\mu}}_{\alpha_0 i}, \hat{\bm{\mu}}_{\alpha_1 i}, ..., \hat{\bm{\mu}}_{\alpha_S i}, \hat{\bm{\mu}}_{\gamma i})$ and simply use the marginal normal distributions of $\hat{\bm{\mu}}_{\alpha_s i}$ for $s=0,...,S$ and that of $\hat{\bm{\mu}}_{\gamma i}$. This decreases the computational burden, and yet produces very similar results compared to the above-described approach. We call this approach a proxy to our original method; details of this appear in Section S.3 of the supplementary materials. A numerical example is also included in the same section of the supplementary materials demonstrating that under the ``small covariance condition" just mentioned,  the confidence intervals by the proxy method are nearly identical to those obtained by the full fiducial approach.

\subsubsection{Fiducial Pivot Statistics for the Fixed Effects}
Following the notations in Section 2.1, we denote the maximum likelihood estimate of the fixed effects vector $\bm{\beta}$ as $\hat{\bm{\beta}}$. Also let $\bm{Y}^* = \begin{pmatrix}
{\bm{Y}_1^*}^t & \hdots & {\bm{Y}_N^*}^t
\end{pmatrix}^t$, $\bm{\mathcal{X}} = \begin{pmatrix}
{\bm{\mathcal{X}}_1}^t & \hdots & {\bm{\mathcal{X}}_N}^t
\end{pmatrix}^t$ and $\bm{\Sigma}_C = \bm{I}_N \otimes \bm{\Sigma}_{Y^*}$. Using the properties of LMM in (3), $\hat{\bm{\beta}} = (\bm{\mathcal{X}}^t{\bm{\Sigma}_C}^{-1}\bm{\mathcal{X}})^{-1}\bm{\mathcal{X}}^t{\bm{\Sigma}_C}^{-1}\bm{Y}^*$ and 
$$
\hat{\bm{\beta}} \sim N_{dL}(\bm{\beta}, (\bm{\mathcal{X}}^t{\bm{\Sigma}_C}^{-1}\bm{\mathcal{X}})^{-1}), \mbox{ asymptotically}.
$$
Let $\bm{Z}_{\beta} = (\bm{\mathcal{X}}^t{\bm{\Sigma}_C}^{-1}\bm{\mathcal{X}})^{\frac{1}{2}}(\hat{\bm{\beta}}-{\bm{\beta}})$, so that $\bm{Z}_{\beta} \sim N_{dL}(\bm{0}_{dL}, \bm{I}_{dL})$, asymptotically. An approximate fiducial pivot statistic for $\bm{\beta}$ can be obtained as
\begin{equation}{\label{ref_beta}}
    \widetilde{\bm{\beta}} = \hat{\bm{\beta}}_o - (\bm{\mathcal{X}}^t{\tilde{\bm{\Sigma}}_C}^{-1}\bm{\mathcal{X}})^{-\frac{1}{2}}\bm{Z}_{\beta},
\end{equation}
where $\hat{\bm{\beta}}_o$ denotes the observed value of $\hat{\bm{\beta}}$ and ${\tilde{\bm{\Sigma}}_C}$ is obtained by replacing the elements of ${\bm{\Sigma}_C}$ with the respective fiducial quantities. Using the expression for $\bm{Z}_{\beta}$ given above and using its asymptotic normal distribution, it is readily verified that $\widetilde{\bm{\beta}}$ in (\ref{ref_beta}) is an approximate fiducial quantity for $\bm{\beta}$. 

\medskip

\noindent {\bf Remark}. In the context of univariate linear mixed models, a methodology is described in  \cite{cisewski2012} for developing fiducial quantities.
However, their method is not directly applicable to our setup since we are dealing with the multivariate mixed effects model (\ref{LMM}). Rather, we have relied on the methodology developed in \cite{krishnamoorthy2007} in order to construct a  fiducial quantity for the covariance matrix of a multivariate normal distribution. We have compared the fiducial approach developed in our work with that in \cite{cisewski2012} in the context of a univariate linear mixed model; please see the simulation study and conclusions reported in Section S.4 in the supplementary material.

\subsection{Fiducial Confidence Limits for  $CCC_L$}
As noted in Section 2, $CCC_L$ does not have any closed-form expression in general, but its value (or its estimate)  can be computed numerically using numerical integration or Monte Carlo integration, given the model parameters (or their estimates).
A fiducial quantity for the $CCC_L$ can be similarly computed by plugging in the fiducial quantities of the corresponding model parameters. Let us denote the fiducial pivot statistics of $\sigma_{\alpha ll\prime}^s$ and $\sigma_{\gamma ll\prime}$ as $\widetilde{\sigma}_{\alpha ll\prime}^s$ and $\widetilde{\sigma}_{\gamma ll\prime}$, respectively, $s=1,...,S$. From (\ref{ref_beta}), the fiducial pivot statistic for the fixed effects vector for rater $l$ is denoted as $\widetilde{\bm{\beta}}_l$ for $l=1,...,L$. Let $\widetilde{{\rm cov}}(\bm{y}_{il}, \bm{y}_{il'})$, $\widetilde{{\rm var}}(\bm{y}_{il})$ and $\widetilde{E}(\bm{y}_{il})$ for $l,l' = 1,...,L$ denote the fiducial quantities for
${{\rm \widetilde{cov}}}(\bm{y}_{il}, \bm{y}_{il'})$, $\widetilde{{\rm var}}(\bm{y}_{il})$ and ${\widetilde{E}}(\bm{y}_{il})$ for $l,l' = 1,...,L$, respectively, obtained by replacing the parameters appearing in these quantities by their respective fiducial quantities. 
The required integrals can be evaluated using one of the aforementioned numerical methods. A fiducial quantity  for $CCC_L$, say $\widetilde{CCC}_L$ can be obtained from its definition  in (\ref{CCCL}):
\begin{equation}{\label{R_CCC_L}}
    \fontsize{11pt}{11pt}\selectfont
    \widetilde{CCC}_L = \frac{2\sum \limits_{l=1}^{L-1} \sum \limits_{l'=l+1}^{L}{\rm tr}(\widetilde{{\rm cov}}(\bm{y}_{il}, \bm{y}_{il'}))}{(L-1)\sum \limits_{l=1}^L {\rm tr}(\widetilde{{\rm var}}(\bm{y}_{il})) + \sum \limits_{l=1}^{L-1} \sum \limits_{l'=l+1}^{L}(\widetilde{E}(\bm{y}_{il}) - \widetilde{E}(\bm{y}_{il'}))^T(\widetilde{E}(\bm{y}_{il}) - \widetilde{E}(\bm{y}_{il'}))}.
\end{equation}

\noindent Under a linear mixed effects model, a closed-form expression is available for $CCC_L$; the expression is denoted by  $CCC_L^{LM}$, and is given in (\ref{CCCL-LM}). Correspondingly, a fiducial quantity is explicitly available for  $CCC_L^{LM}$, and is given by 
\begin{equation}{\label{R_CCC_LM}}
\fontsize{9pt}{9pt}\selectfont
     \widetilde{CCC}_L^{LM} =\frac{\sum \limits_{l = 1}^{L-1}\sum \limits_{l^\prime = l+1}^{L} \left[\sum \limits_{s=0}^{S} \widetilde{\sigma}_{\alpha l l^\prime}^s \bm{z_s}^t\bm{z_s} + KT \widetilde{\sigma}_{\gamma l l^\prime} \right]}{(L-1)\sum \limits_{l=1}^L \left[\sum \limits_{s=0}^{S} \widetilde{\sigma}_{\alpha ll}^s \bm{z_s}^t\bm{z_s} + KT \widetilde{\sigma}_{\gamma ll} + K \sum \limits_{j=1}^T \widetilde{\sigma}^2 \right] + \sum \limits_{l = 1}^{L-1}\sum \limits_{l^\prime = l+1}^{L} \left[ \frac{1}{N} \sum \limits_{i=1}^N (\widetilde{\bm{\beta}}_l - \widetilde{\bm{\beta}}_{l^\prime})^t \bm{X}_i^t \bm{X}_i (\widetilde{\bm{\beta}}_l - \widetilde{\bm{\beta}}_{l^\prime}) \right]},
\end{equation}
\noindent where, $\widetilde{\sigma}^2$ is a fiducial quantity for $\sigma^2$. 

Given the observed data, we can obtain fiducial confidence limits based on the distribution of $\widetilde{CCC}_L$.  In order to generate a realization of $\widetilde{CCC}_L$, we need to: (i) generate realizations of independent standard normal and chi-squares random variables, and (ii) solve a non-linear least squares optimization problem or a system of non-linear equations. To find a $100(1-\alpha)\%$ confidence interval of $CCC_L$, $10,000$ independent samples are generated for $\widetilde{CCC}_L$ and the highest density region covering $100(1-\alpha)\%$ of the total area under the simulated density  is determined following the algorithm in \cite{kruschke2014}. We denote the vector of parameters of the model by $\bm{\Theta} = \{\phi, \sigma_{\alpha ll\prime}^s, \sigma_{\gamma ll\prime}, \bm{\beta}_l), s=0,1,...,S; l,l^{'}=1,...,L\}$, and the fiducial  quantity for  $\bm{\Theta}$ is denoted by $\widetilde{\bm{\Theta}} = \{(\widetilde{\phi}, \widetilde{\sigma}_{\alpha ll\prime}^s, \widetilde{\sigma}_{\gamma ll\prime}, \widetilde{\bm{\beta}}_l), s=0,1,...,S; l,l^{'}=1,...,L\}$, where the procedure for construction of $\widetilde{\bm{\Theta}} $ is described above. We are now ready to put our results in the form of a theorem.
\medskip

\noindent{\it Theorem 2}. Let  $\bm{\Theta}$ be the vector of model parameters and   $\widetilde{\bm{\Theta}}$  a fiducial quantity for  $\bm{\Theta}$ so that (i) given the observed data $\bm{y}$, the distribution of $\widetilde{\bm{\Theta}}$ is free of any unknown parameters, and (ii) if the random variables in $\widetilde{\bm{\Theta}}$ are replaced by the corresponding observed data, $\widetilde{\bm{\Theta}}$ simplifies to $\bm{\Theta}$.  Then $\widetilde{CCC}_L$ given in (\ref{R_CCC_L}), a function of the components of $\widetilde{\bm{\Theta}}$, is a fiducial quantity for  ${CCC}_L$.

\noindent{\it Proof}. The theorem follows by noting that (i) if $h({\bm{\Theta}})$ is a function of ${\bm{\Theta}}$, then  $h(\widetilde{\bm{\Theta}})$ is a fiducial quantity for  $h({\bm{\Theta}})$, and (ii) the CCC is a function of ${\bm{\Theta}}$.

\section{Simulation Study}
In this section, we shall report the results from a simulation study in order to assess the performance of our proposed fiducial confidence intervals for the CCC under two models: (i) a Gaussian family corresponding to a linear mixed effects model, and (ii) the scenario of longitudinal ratings on the count scale, where a Poisson distribution with
a log-link function is appropriate. We have also compared the fiducial confidence intervals to those based on the Fisher's Z transformation and those based on the bootstrap. 

\subsection{Gaussian family: Linear mixed effects model}
Under a linear mixed-effects model (LMM), $h(.)$  in (\ref{mean}) is the identity function. Hence $\bm{\mu} = \bm{\eta}$ and $\mathcal{G}^{\prime}(.) = \bm{I}$ which jointly imply $\bm{y} = \bm{y}^*$ in (3). Additionally, $\zeta(.) = 1$ and the dispersion parameter $\phi = \sigma^2$, which is the error variance. If $y_{ijkl}$ denotes the $k$th rating for the $i$th subject at the $j$th time point by rater $l$, the model used for the simulation is 
\begin{equation}{\label{example_LMM}}
    y_{ijkl} =  \beta^{(l)}_{0}+ \beta^{(l)}_{1}j + \alpha^{(l)}_{0i} + \alpha^{(l)}_{1i}j + \gamma^{(l)}_{0ij} + \epsilon_{ijkl},
\end{equation}
$i=1,...,N$, $j=1,...,T$, $k=1,...,K$, $l=1,...,L$. In terms of the notation used in (\ref{mean}) and (\ref{LMM}), we now have $\bm{X}_i = \begin{pmatrix} \bm{1}_{KT} & \bm{1}_K 
\otimes \bm{t} \end{pmatrix}$, for $\bm{t} = (1 \hdots T)^t$, $\bm{\beta}_l = \begin{pmatrix} \beta^{(l)}_0 & \beta^{(l)}_1 \end{pmatrix}^t$; $S=1$ with $\bm{z}_0 = \bm{1}_{KT}$, $\bm{z}_1 = \bm{1}_K 
\otimes \bm{t}$, $\alpha^0_{i1} = \alpha^{(l)}_{0i}$, $\alpha^1_{i1} = \alpha^{(l)}_{1i}$ and $\gamma_{ijl} = \gamma^{(l)}_{0ij}$. The model assumptions stated for (\ref{mean}) and (\ref{LMM}) are similarly applicable for (\ref{example_LMM}).
The expression of the CCC based on the model in (\ref{example_LMM}) simplifies to
\begin{equation}{\label{CCC_example_LM}}
    \fontsize{9pt}{14pt}\selectfont
    CCC^{G}_L = \frac{\sum \limits_{l = 1}^{L-1}\sum \limits_{l^\prime = l+1}^{L} \left[T(\sigma_{\alpha ll^{\prime}}^0 + \sigma_{\gamma ll^{\prime}}) + \sum \limits_{j=1}^T \sigma_{\alpha ll^{\prime}}^1j^2 \right]}{(L-1)\sum \limits_{l=1}^L \left[T(\sigma_{\alpha ll}^0 + \sigma_{\gamma ll}) + \sum \limits_{j=1}^T \sigma_{\alpha ll}^1j^2 + T \sigma^2 \right] + \sum \limits_{l = 1}^{L-1}\sum \limits_{l^\prime = l+1}^{L} \sum \limits_{j=1}^T\left[(\beta^{(l)}_0 - \beta^{(l^\prime)}_0) + (\beta^{(l)}_1 - \beta^{(l^\prime)}_1)j\right]^2}.
\end{equation}
\noindent We shall first briefly explain the computation of a fiducial quantity  for the dispersion parameter  $\sigma^2$. Fiducial quantities  for the other parameters can then be formed using the methods discussed in Section 3.

We first note that the total number of fixed effects and random effects in the model  (\ref{example_LMM}) is $2L$ (number of fixed effects) + $2NL$ (number of subject level random effects) + $NTL$ (number of subject-time level random effects).  We also have a total of $NTKL$ observations. If $\hat{\sigma}^2$  denotes the MLE of the dispersion parameter $\sigma^2$, then we shall assume the distribution 
$$
(NTKL - 2L - 2NL - NTL)\hat{\sigma}^2 \widesim[1]{a} \sigma^2\chi^2_{(NTKL - 2L - 2NL - NTL)}.
$$
We note that in a linear mixed effects model with the usual normality assumptions, if $\hat{\sigma}^2$ is the unbiased estimator of $\sigma^2$ based on the error sum of squares in the ANOVA decomposition, then the above chi-squares distribution is exact. We shall proceed under the assumption that the above distributional assumption is valid at least approximately.  If $\hat{\sigma}^2_o$ denotes the observed value of $\hat{\sigma}^2$, a fiducial quantity for $\sigma^2$ is then given by 
$(NTKL - 2L - 2NL - NTL)\hat{\sigma}^2_o/U$ where $U \sim \chi^2_{(NTKL - 2L - 2NL - NTL)}$. This fiducial quantity for $\sigma^2$ is available in the literature; see,  \cite{mathew_book} (Section 1.4.3).

 We shall now report numerical results in order to assess the performance of the proposed fiducial confidence interval in terms of both coverage probabilities and expected widths under different simulation settings. An algorithm for computing the coverage probability is given in Section S.2 of the supplementary material. The proposed fiducial approach will also be compared with the  following two alternative approaches:\\
\textbf{Bias-Corrected Parametric Bootstrap}: The parametric bootstrap approach is based on parametric bootstrap samples generated based on the Model (\ref{example_LMM}) using estimated parameters. The CCC is evaluated from each parametric bootstrap sample and the confidence interval is obtained from the bootstrap based empirical quantiles. Furthermore, a bias correction has also been applied; see \cite{hastiebook} (Chapter 10).\\
\textbf{Fisher Z Transformation}: \cite{lin1989} gave an expression for the variance of Fisher Z transformed CCC when the ratings are continuous and jointly follow a bivariate normal distribution. This large sample-based approach is often used by most practitioners for the interval estimation of the CCC, even for non-normal data with moderate or small sample sizes. In our numerical results, we have included this confidence interval too.

Table \ref{tab_gaussian} gives the coverage probabilities and expected widths of the 95\% confidence intervals for the CCC in the set up of the model (\ref{CCC_example_LM}) for a two-rater design with $K$ = 1, assuming that the subject-time random interaction effects are absent. 
Also, for the results in Table \ref{tab_gaussian}, we assumed that each subject was observed over 10 time points, i.e. $T=10$. The values of the other relevant model parameters used in the simulations are given in Table \ref{tab_gaussian}.  From the numerical results in Table \ref{tab_gaussian} we see that the confidence interval based on Fisher's Z transformation is quite conservative, resulting in a comparatively wider interval even for a sample size of $N$ = 50. The interval based on the parametric bootstrap falls short in terms of meeting the coverage probability requirement, even for the sample size  $N$ = 50. The performance of the proposed fiducial interval is significantly better in terms of both coverage probability and expected width.     
In fact, the fiducial approach produces a significantly narrower confidence interval compared to the other two competing methods. For smaller sample sizes ($N$ = 15 in Table \ref{tab_gaussian}), the coverage probability of the fiducial interval is slightly below the nominal level of 95\%. However,  as the sample size becomes slightly larger, the fiducial approach gradually achieves the target coverage probability maintaining narrower expected widths compared to Fisher Z and bootstrap.  Additional simulated results are given in Table S.1 in the Supplementary File for a mixture of normal and gamma and another mixture of normal and lognormal distritubions. Better performance of the fiducial approach compared to Fisher Z and bootstrap  in terms of the expected width and coverage  probability are noted for sample sizes $N$ = 30, 50 and 100.

We did a limited numerical investigation of the robustness of the fiducial approach by considering some mixture distributions, and by adding outliers from highly skewed and heavy-tailed distributions such as Gamma and Log-normal to the normal errors. The relevant numerical results appear in the Supplementary File.  Under the mixture scenario just mentioned, we observe from Table S.1  that the simulated coverage probability of the fiducial approach is very close to the target value compared to the other two competitors. In general, Fisher Z is extremely conservative. Coverage probabilities for both Fisher Z and the parametric bootstrap fluctuated more when the intensity of skewness increased. As mentioned in Section S.2 in the Supplementary File, the coverage probabilities presented in all  tables have been calculated based on 10000 simulated datasets.

\begin{table}[h]
\caption{\textbf{Estimated coverage probabilities and expected widths of the 95\% confidence intervals based on the fiducial approach, Fisher Z transformation and bias-corrected parametric bootstrap under a two-level design for two raters with $K=1$, $T$ = 10  and no subject-time interaction.  Bounds for the CCC  are computed as  -0.961 and 0.961 (using Theorem 1 )}} 
\vspace{.4cm}
\renewcommand{\arraystretch}{1}
\fontsize{9}{10}\selectfont
    \centering
     \begin{tabular}{|c|c|c|c|c|c|c|}
    \hline
     & CCC & Sample & Method & Expected & Expected & Coverage\\
    Parameters & (True) & Size & & Confidence & Width & Probability\\
    &        & (N) & & Limits & & \\
     \hline
     \multirow{9}{*}{\fontsize{7}{10}\selectfont
  \begin{tabular}{c}
        $\beta_0^{(1)} = 0.75$, $\beta_0^{(2)} = 0.50$\\\vspace{.3cm}
        $\beta_1^{(1)} = -0.10$, $\beta_1^{(2)} = -0.06$\\
        $\begin{pmatrix} \sigma^0_{\alpha11} & \sigma^0_{\alpha12}\\
        \sigma^0_{\alpha12} & \sigma^0_{\alpha22} \end{pmatrix} = 
        \begin{pmatrix} 0.45 & 0.40\\
        0.40 & 0.49 \end{pmatrix}$\\\vspace{.3cm}
        $\begin{pmatrix} \sigma^1_{\alpha11} & \sigma^1_{\alpha12}\\
        \sigma^1_{\alpha12} & \sigma^1_{\alpha22} \end{pmatrix} = 
        \begin{pmatrix} 0.10 & 0.067\\
        0.067 & 0.06 \end{pmatrix}$\\
        $\sigma^2 = 0.11$
  \end{tabular}} & \multirow{9}{*}{0.805} & \multirow{3}{*}{15} & Fiducial & (0.606, 0.890) & \textbf{0.284} & \textbf{0.939}\\
   & & & Fisher Z transform & (0.491, 0.913) & 0.422 & 0.996\\
   & & & Bootstrap & (0.586, 0.897) & 0.311 & 0.905\\
     \cline{3-7}
    & & \multirow{3}{*}{30} & Fiducial & (0.688, 0.870) & \textbf{0.192} & \textbf{0.942}\\
   & & & Fisher Z transform & (0.628, 0.893) & 0.275 & 0.992\\
   & & & Bootstrap & (0.661, 0.895) & 0.213 & 0.921\\
     \cline{3-7}
    & & \multirow{3}{*}{50} & Fiducial & (0.722, 0.858) & \textbf{0.136} & \textbf{0.946}\\
    & & & Fisher Z transform & (0.679, 0.877) & 0.202 & 0.998\\
    & & & Bootstrap & (0.718, 0.864) & 0.146 & 0.935\\
     \hline
    \end{tabular}
    \label{tab_gaussian}
\end{table}

\subsection{Poisson Family}
When longitudinal ratings are on the count scale, a Poisson distribution with a log-link function can generally  be used to model the data.   Taking $h(.)$ in Model (\ref{mean}) to be the \textit{exponential} function, we express $\bm{\mu} = exp(\bm{\eta})$ and $\mathcal{G}^{\prime}(\bm{\mu}) = [diag(exp(\bm{\eta}))]^{-1} = diag(\bm{\mu})^{-1}$, implying $\bm{y}^* = \hat{\bm{\eta}} + diag(\hat{\bm{\mu}})^{-1}(\bm{y} - \hat{\bm{\mu}})$ in (\ref{LMM}).  We also consider $\zeta(\mu) = \mu$ and the dispersion parameter $\phi = 1$. Based on (\ref{approx}), conditional on the random effects, var$(\bm{\epsilon})$ = diag$(\bm{\mu})$, and this implies, $\bm{\epsilon}^* \sim N(\bm{0}, diag(\bm{\mu})^{-1})$ in (\ref{LMM}). For  simulation, instead of referring to the general model in (\ref{mean}), we have used a very specific model for rater $l$, described below. 

\begin{equation}{\label{example_poisson}}
    E(y_{ijkl}|\alpha^{(l)}_{0i}, \alpha^{(l)}_{1i}, \gamma^{(l)}_{0ij})= \mu_{ijkl} = exp\left(\beta^{(l)}_{0}+ \beta^{(l)}_{1}j + \alpha^{(l)}_{0i} + \alpha^{(l)}_{1i}j + \gamma^{(l)}_{0ij}\right), \quad l=1,...,L,
\end{equation}
After transforming $y_{ijkl}$ to $y_{ijkl}^* = log(\hat{\mu}_{ijkl}) + \frac{1}{\hat{\mu}_{ijkl}}(y_{ijkl} - \hat{\mu}_{ijkl})$, the transformed model in (\ref{LMM}) can now be simplified as,
\begin{equation}{\label{example_poisson_LMM}}
    y_{ijkl}^*= \beta^{(l)}_{0}+ \beta^{(l)}_{1}j + \alpha^{(l)}_{0i} + \alpha^{(l)}_{1i}j + \gamma^{(l)}_{0ij} + \epsilon_{ijkl}^*,
\end{equation}
where $\epsilon_{ijkl}^*|\alpha^{(l)}_{0i}, \alpha^{(l)}_{1i}, \gamma^{(l)}_{0ij} \widesim[1.5]{indp} N(0, \frac{1}{\mu_{ijkl}})$ for $i=1,..., N$, $j=1,...,T$, $k=1,...,K$, $l=1,...,L$. In terms of the notations used in (\ref{mean}) and (\ref{LMM}), we now have $\bm{X}_i = \begin{pmatrix} \bm{1}_{KT} & \bm{1}_K 
\otimes \bm{t} \end{pmatrix}$, for $\bm{t} = (1 \hdots T)^t$, $\bm{\beta}_l = \begin{pmatrix} \beta^{(l)}_0\\ \beta^{(l)}_1 \end{pmatrix}$; $S=1$ with $\bm{z}_0 = \bm{1}_{KT}$, $\bm{z}_1 = \bm{1}_K 
\otimes \bm{t}$, $\alpha^0_{i1} = \alpha^{(l)}_{0i}$, $\alpha^1_{i1} = \alpha^{(l)}_{1i}$ and $\gamma_{ijl} = \gamma^{(l)}_{0ij}$. We are omitting the model assumptions here as they are exactly the same as those for the model (\ref{mean}) and the model (\ref{approx}).

 For the Poisson family, an analytical expression of $CCC_L$ can be computed with a closed form as the integrals involved in computing expectations, variances, and covariances are the same as the moment-generating function of a multivariate normal distribution. The expression of $CCC_L$ based on the model in (\ref{example_poisson}) can be simplified as
\begin{equation}{\label{CCC_example_poisson}}
    \fontsize{9pt}{11pt}\selectfont
    CCC^P_L = \frac{\sum \limits_{l = 1}^{L-1}\sum \limits_{l^\prime = l+1}^{L} \left[\sum \limits_{j=1}^T \lambda_{lj} \lambda_{l'j} (exp(\sigma_{\alpha ll'}^0 + \sigma_{\gamma ll'} + \sigma_{\alpha ll^{\prime}}^1j^2) - 1) \right]}{(L-1)\sum \limits_{l=1}^L \left[\sum \limits_{j=1}^T \lambda_{lj}\{\lambda_{lj} (exp({\sigma_{\alpha ll}^0}^2 + {\sigma_{\gamma ll}}^2 + {\sigma_{\alpha ll}^1}^2j^2) - 1) + 1\} \right] + \sum \limits_{l = 1}^{L-1}\sum \limits_{l^\prime = l+1}^{L} \sum \limits_{j=1}^T\left[(\lambda_{lj} - \lambda_{l'j})^2\right]},
\end{equation}
\noindent where, $\lambda_{lj} = exp(\beta_0 + \beta_1j + \frac{1}{2}(\sigma_{\alpha ll}^0 + \sigma_{\gamma ll} + \sigma_{\alpha ll}^1j^2))$. 

In Table \ref{tab_poisson}, we compare the performance of the following two approaches along with the large sample based Fisher Z approach of \cite{lin1989} (described in Section 4.1) for evaluating the confidence interval of CCC for the Poisson family. Note that, in the Fisher Z approach, we have used the closed-form expression in (\ref{CCC_example_poisson}) to compute the CCC.\\
\textbf{Fiducial with Exact CCC}: This is the proposed fiducial approach (in Section 3) which uses the closed-form expression given in (\ref{CCC_example_poisson}) to compute a fiducial quantity for the CCC, i.e., the fiducial quantity for the CCC is obtained  by replacing the fixed effects and variance components in (\ref{CCC_example_poisson}) by their respective fiducial quantities. The fiducial confidence interval so obtained is termed  ``Fiducial with Exact CCC".\\
\textbf{Fiducial with Numerical CCC}: The fiducial quantity obtained by this approach is essentially the same as the one just obtained,  except in the way the computation is carried out. As discussed in Section 2, when no analytical expression is available for the  CCC, we numerical approximation is used for the integrals in the expression of the CCC. Thus, similar to the previous approach, we first develop fiducial statistics for all the model parameters. Next, utilizing these fiducial statistics, we implement Monte Carlo integration for computing the expectation, variance, and covariance terms given in the expression of the fiducial pivot statistic for the CCC, given in (\ref{R_CCC_L}). The fiducial confidence interval of the CCC so obtained is termed as  ``Fiducial with Numerical CCC". Note that, for the Poisson GLMM the exact expression of the CCC is available in  (\ref{CCC_example_poisson}). Thus, the numerical approach to obtain CCC is not necessary; however, this provides an opportunity to assess the performance of the fiducial approach when a closed-form expression for the CCC is not available, which is the case for several GLMMs. We compare the performance of the ``Fiducial with Exact CCC" with that of the ``Fiducial with Numerical CCC" via simulations and the results are given in Table \ref{tab_poisson}.

Table \ref{tab_poisson} displays the performance of the aforementioned approaches. A three-level design with two raters has been used for this simulation. The parameters used for the simulations are: $\beta_0^{(1)} = 4.50$, $\beta_0^{(2)} = 4.30$, $\beta_1^{(1)} = -0.03$, $\beta_1^{(2)} = 0.03$, $\begin{pmatrix} \sigma^0_{\alpha11} & \sigma^0_{\alpha12}\\
\sigma^0_{\alpha12} & \sigma^0_{\alpha22}
\end{pmatrix} = \begin{pmatrix} 0.63 & 0.60\\ 0.60 & 0.66 \end{pmatrix}$, $\begin{pmatrix} \sigma^1_{\alpha11} & \sigma^1_{\alpha12}\\
\sigma^1_{\alpha12} & \sigma^1_{\alpha22}
\end{pmatrix} = \begin{pmatrix} 0.08 & 0.05\\ 0.05 & 0.07 \end{pmatrix}$. Unlike the Gaussian family, the Poisson mixed-effects regression model involves an additional approximation for linearization via equation (16).
Perhaps this explains the slight underperformance of the fiducial approach even with Exact CCC for smaller sample sizes such as $n$ = 15  in terms   of the  expected width and the coverage  probability, compared to the fiducial approach used in linear mixed effects model discussed in the previous section (Section 4.1). In general, performance of the ``Fiducial with Numerical CCC" is very similar to that of ``Fiducial with Numerical CCC". However the  extreme conservatism of the Fisher Z is evident everywhere.

\begin{table}[h]
\caption{\textbf{A comparison among Exact CCC, Numerical CCC and Fisher Z for Poisson Family.  Each subject is observed over $10$ time points and at each time point $5$ observations are taken by each rater, i.e. $T=10$, $K=5$.} }
\centering
    \vspace{.4cm}
    \renewcommand{\arraystretch}{1.2}
\fontsize{10}{10}\selectfont
    \centering
     \begin{tabular}{|c|c|c|c|c|c|}
    \hline
    CCC & Sample & Method & Average 95\% & Expected Width & Inclusion\\
    (True) & Size & & Confidence & of 95\% & Probability\\
            & (N) & & Limits & Confidence Limits & \\
     \hline
     
    \multirow{12}{*}{0.822} & \multirow{3}{*}{15} & Fiducial with Exact CCC & (0.568, 0.883) & 0.315 & 0.929\\
    & & Fiducial with Numerical CCC & (0.567, 0.879) & 0.312 & 0.925\\
    & & Fisher Z Transformation & (0.524, 0.921) & 0.397 & 0.996\\
    \cline{2-6}
    & \multirow{3}{*}{30} & Fiducial with Exact CCC & (0.687, 0.873) & 0.186 & 0.954\\
    & & Fiducial with Numerical CCC & (0.681, 0.873) & 0.192 & 0.960\\
    & & Fisher Z Transformation & (0.656, 0.910) & 0.254 & 0.999\\
    \cline{2-6}
    &  \multirow{3}{*}{50} & Fiducial with Exact CCC & (0.729, 0.864) & 0.135 & 0.952\\
    & & Fiducial with Numerical CCC & (0.721, 0.862) & 0.141 & 0.956\\
    & & Fisher Z Transformation & (0.706, 0.893) & 0.245 & 0.997\\
    \cline{2-6}
    &  \multirow{3}{*}{100} & Fiducial with Exact CCC & (0.765, 0.855) & 0.090 & 0.950\\
    & & Fiducial with Numerical CCC & (0.764, 0.855) & 0.091 & 0.951\\
    & & Fisher Z Transformation & (0.753, 0.880) & 0.127 & 0.996\\
     \hline
    \end{tabular}
    \label{tab_poisson}
\end{table}
Further, using Theorem 1 the bounds of CCC are computed as -0.992 and .992.

\section{ Data Analysis for the Examples}
We now discuss the data analysis corresponding to the motivational examples described in Section 1.1. We have used the GLMM described in (\ref{mean}) for each example. Further, for the computation of confidence intervals for the CCC among two or more raters, we have used the proposed fiducial approach, and also the approach based on the Fisher's Z transformation. We did not implement the parametric bootstrap approach in view of its poor performance in terms of the coverage probability. Finally, we conclude with a contextual interpretation based on the computed confidence intervals.

In the GAIT study (Example 1), the radiographs from each of $38$ subjects with complete observations  were read by each of two physician investigators and the non-technician (by computer) independently at $3$ different time points, and each reading was repeated $2$ times.  Following earlier notation, we have: $L=3$, $N=38$, $T=3$, and $K=2$. The readings by the raters were continuous, but we observed the data to be a slightly right-skewed, which motivated us to use a 3-level Gamma GLMM with an \textit{inverse}-link function. In this three-raters scenario, we have calculated confidence intervals for the CCC for each of the three pairs as well as for the CCC among all the three raters. The results are given in Table \ref{tab_data}.

In the second example, $10$ healthy subjects were enrolled, and for each subject, the measurements of fiber number in both the right and left corticospinal tracts were taken by a well-trained technologist and a neuroradiologist. In addition, two scans in one session and a third scan one week later were collected. Further, the bilateral corticospinal tracts in each scan were reconstructed twice. Therefore, we have $L=2$, $n = 10$, $T = 3$, and $K = 2$ in the study. For the corresponding count data, we used a Poisson GLMM with log-link function. Results are once again given in Table \ref{tab_data}.

Inspecting Table \ref{tab_data}, we note that the difference between the confidence intervals based on the fiducial approach and those based on the Fisher's Z transformation is consistent with the conclusions from the numerical results in the previous section. In particular, the Fisher's Z transformation based confidence intervals are significantly wider compared to the corresponding fiducial interval. Focusing on the fiducial confidence intervals, we can assert with high confidence that the CCC's under consideration are all more than 0.5 for both the examples. The overall conclusion is that all the CCC's appear to be somewhat high, indicating significant agreement among the raters. The examples and the numerical results give strong justifications for using the fiducial approach.  

\begin{table}[h]
\caption{\textbf{Confidence intervals under a Gamma GLMM with the inverse-link for Example 1 and a Poisson GLMM with the log-link for Example 2. CCC limits of  Example 1, Physician 1 vs Physician 2: (-0.952, 0.952), Physician 1 vs Computer: (-0.947, 0.947), Physician 2 vs Computer: (-0.961, 0.961), Three raters combined: (-0.918, 0.918). For Example 2, the limits are (-0.995, 0.995).}}
    \renewcommand{\arraystretch}{1}
    \fontsize{9}{10}\selectfont
    \centering
     \begin{tabular}{|c|c|c|c|c|c|}
    \hline
    Data & Model & CCC Type & Method & 95\% Confidence & CI Width\\
    & & & & Interval (CI) &\\
     \hline
     
    \multirow{8}{*}{Example 1} & \multirow{8}{*}{Gamma GLMM} & \multirow{2}{*}{Physician 1 vs Physician 2} &  Fiducial  & (0.675, 0.864) & 0.189\\
    & & & Fisher Z Transform & (0.618, 0.892) & 0.286\\
    \cline{3-6}
    & & \multirow{2}{*}{Physician 1 vs Computer} & Fiducial & (0.553, 0.771) & 0.218\\
     & & & Fisher Z Transform & (0.492, 0.807) & 0.315\\
     \cline{3-6}
    & & \multirow{2}{*}{Physician 2 vs Computer} & Fiducial & (0.608, 0.817) & 0.209\\
     & & & Fisher Z Transform & (0.538, 0.841) & 0.303\\
     \cline{3-6}
    & & \multirow{2}{*}{Three raters combined} & Fiducial & (0.582, 0.831) & 0.249\\
     & & & Fisher Z Transform & (0.492, 0.879) & 0.387\\
    \hline
    \multirow{2}{*}{Example 2} & \multirow{2}{*}{Poisson GLMM} & \multirow{2}{*}{Technologist and Neuroradiologist} & Fiducial & (0.664, 0.940) & 0.276\\
     & & & Fisher Z Transform & (0.516, 0.960) & 0.444\\
     \hline
    \end{tabular}
    \label{tab_data}
\end{table}

\section{Discussion}

Since the fundamental work of \cite{lin1989}, the topic of assessing agreement among different raters or methods has received considerable attention in the literature, leading to several publications and two books: \cite{linbook}  and \cite{choudharybook}. In particular, the concordance correlation coefficient (CCC), originally proposed in \cite{lin1989}, has been extended to various scenarios in order to accommodate the structure of the data encountered in applications where measuring agreement is of paramount concern. Clearly, the structure of the data (discrete or continuous, the longitudinal nature of the data, etc.) and the models for the data, have to be taken into account while defining the CCC. In addition, the computation of confidence intervals for the CCC, that are satisfactory from the perspective of maintaining the coverage probability, is of obvious interest. This article is an attempt to address the interval estimation problem in a general scenario of a generalized linear mixed model relevant to many applications that call for the assessment of agreement using the CCC. Many standard models (for example, the linear mixed effects model) are special cases of the set up considered in the present work.  

Since the Fisher’s Z-transformation is a well known and widely used methodology for inference on the usual correlation coefficient, it is quite natural to adopt this for the CCC as well. Another obvious strategy is to rely upon a parametric bootstrap approach for the interval estimation. Our numerical results demonstrate that the former can be very conservative even for large sample sizes, and the latter falls short in terms of the coverage probability. In view of these observations, we have explored a fiducial approach for the interval estimation of the CCC, based on a linearization of the model, already available in the literature. For reasonable sample sizes,  the fiducial approach has resulted in confidence intervals that  accurately maintain the coverage probability, even though the latter can be below the nominal level in small sample size scenarios. An important observation is that when the coverage probabilities are close to the nominal level, the fiducial approach provides confidence intervals having a significantly smaller expected width. In the small sample settings where our fiducial approach falls short in terms of the coverage probability, perhaps one can attempt a bootstrap calibration (Chapter 18, \cite{hastiebook}  in order to improve the coverage probability performance, at the cost of an increased computational burden. This is currently under investigation.   

Our approach has the potential for extensive applications within artificial intelligence (AI).   The integration of AI into medical sciences is becoming increasingly prevalent, especially since  the use of AI software and devices are becoming more prominent in clinical diagnosis. 
For example, (\cite{Zelt2023}) have investigated agreement between AI diagnoses with those made by virtual care providers and blind adjudicators in the context of virtual primary care. For diagnosing glaucoma using the vertical cup to disc ratio (VCDR) measurements,  (\cite{Shro2023})  have discussed the assessment of  the agreement of VCDR measured by a new AI software, with those obtained by spectral-domain optical coherence tomography , and manual grading by experts. In the context of routine breast cancer diagnostics, (\cite{Nicl2023}) have examined agreement rates of pathologist scores with and without AI assistance. In the article (\cite{Nicl2023}) cited earlier, addressing agreement rates of pathologist scores for routine breast cancer diagnosis, with and without AI assistance,  the authors express the concern that  “faulty AI analysis may bias the pathologist
and contribute to incorrect diagnoses and, therefore, may lead to inappropriate therapy or prognosis.”

Despite the widespread adoption of AI in healthcare, only a limited number of AI systems are routinely utilized in clinical settings, primarily due to a lack of reliability studies. Current AI analyses may result in incorrect diagnoses and inappropriate prognoses, highlighting the critical need for rigorous evaluation and validation of AI technologies before their widespread clinical implementation.  This prompted the authors to look at agreement rates of pathologist scores with and without AI assistance in order to draw conclusions on the reliability of the AI tool. An equally important issue is whether performances of multiple experts are in  agreement among themselves before singling out AI devices. A more interesting question is if an expert is asked to evaluate the same task repeatedly over a time gap, whether the evaluations will agree. This brings the notion of intra and inter-rater agreement among human experts that may have a role in the evaluation process of AI devices. The proposed fiducial approach can be utilized to address all such questions.

Another potential application of our approach is in quality control. Medical research that involve  measurements from images requires precision, and the ability to replicate those measurements.  Thus, a common quality control measure is necessary when multiple raters take measurements independently, which is what the CCC accomplishes. We believe that our work provides a unified framework for assessing agreement using the CCC and its interval estimator in a very general modeling framework.

 \section{Summary of the Supplementary File}

We include a Supplementary File with this manuscript, which comprises several key components. First, we outline a methodology for constructing the fiducial quantity of the covariance matrix of a multivariate normal distribution, and the derivations are presented in the Appendix Section. Following this, in Section S.1, we present an algorithm for constructing fiducial confidence intervals for CCC, and in Section S.2, an additional algorithm is provided for obtaining simulated coverage probabilities of CCC. In Section S.3, we detail a proxy approach to determine pivot statistics for the covariance matrices of random effects when the estimated covariance among the random effect predictors is low. In Section S.4, through a simulation study, we have compared the fiducial approach developed in our work with that in \cite{cisewski2012} in the context of a univariate linear mixed model. In Section S.5, we demonstrate the superiority of fiducial methods over Fisher-Z, and bootstrap approaches for a mixture of normal and gamma distributions through simulations, as summarized in Table S.1. Moreover, we present additional results showcasing the performance of fiducial methods in various scenarios. Specifically, Tables S.2 and S.3 depict the fiducial performance for a two-level design with three raters and a three-level design with two raters for the Gaussian Family, respectively. Section S.6 demonstrates another example of the proposed fiducial method with a three-level design with a gamma distribution. These comprehensive analyses not only highlight the efficacy of fiducial methods but also provide valuable insights into their performance across different experimental setups and distributions.

\bibliographystyle{agsm2}
\bibliography{references}


\end{document}


\def\spacingset#1{\renewcommand{\baselinestretch}%
{#1}\small\normalsize} \spacingset{1}

\spacingset{1.9}

\renewcommand\thesubsection{A.\arabic{subsection}}
\section{Appendix}

\subsection{A Fiducial Quantity  for the Covariance Matrix of a Multivariate Normal Distribution}
The fiducial quantity that we shall exhibit, say  $\tilde{\bm{\Sigma}}$, will be a function of the observed data, and pivot statistics based on random samples, such that (i) given the observed data, the distribution of $\tilde{\bm{\Sigma}}$ is free of any unknown parameters, and (ii) if the random variables in the pivot statistics are replaced by the corresponding observed data, $\tilde{\bm \Sigma}$ simplifies to $\bm\Sigma$. 
\noindent Without any loss of generality, we shall focus on a Wishart matrix. Thus, let
$\bm{S} \sim \bm{W}_p(\bm{\Sigma}, n)$, a p-variate Wishart distribution with the scale matrix $\bm{\Sigma}$ and degrees of freedom $n$. Following the idea of \cite{krishnamoorthy2007}, we outline the construction of the fiducial pivot quantity for $\bm{\Sigma}$ below.

Let $\bm{T}_s$ and $\bm{\theta}$ be Cholesky factors of $\bm{S}$ and $\bm{\Sigma}$, respectively, so that $\bm{S} = \bm{T}_s\bm{T}_s^t$, and $\bm{\Sigma} = \bm{\theta}\bm{\theta}^t$. By construction, $\bm{T}_s$ and $\bm{\theta}$ are lower triangular matrices with positive diagonal elements.  Let, $\bm{V} = \bm{\theta}^{-1}\bm{T}_s$, a lower triangular matrix with elements $V_{ij}$, $i=1,...,p$, $j=1,...,p$, $i\geq j$.
Note that 
$$\bm{V}\bm{V}^t = \bm{\theta}^{-1}\bm{T}_s\bm{T}_s^t{\bm{(\theta}}^{-1})^t \sim \bm{W}_p(\bm{I}_p, n).
$$
Thus, the $V_{ij}$'s are all independent with 
$$V_{ii}^2 \sim \chi^2_{n-i+1} \quad i=1,...,p, \quad \quad \quad V_{ij} \sim N(0, 1), \quad i>j,$$
where $\chi^2_m$ denotes the central chi-square distribution with $m$ degrees of freedom.
Let $\bm{t}_s$ be the observed value of the Cholesky factor $\bm{T}_s$. Then  $\bm{R} = \bm{t}_s\bm{V}^{-1}$ is a fiducial quantity for $\bm{\theta}$. Note that $\bm{R}$ is also a lower triangular matrix. A fiducial quantity for $\bm{\Sigma}$ is now given by $\tilde{ \bm{\Sigma}} = \bm{R}\bm{R}^t$. It should be clear that values of $\tilde{\bm{\Sigma}}$ can be generated after generating independent chisquare random variables $V_{ii}^2$, and standard normal random variables  $V_{ij}$, $i>j$ in order to obtain the lower triangular matrix $\bm V$, and then computing  $\tilde{\bm{\Sigma}} = \bm{t}_s(\bm{V}^t\bm{V})^{-1}\bm{t}_s^t$. We note that a fiducial quantity for a scalar valued function $h(\bm\Sigma)$ is given by 
$h(\tilde{\bm{\Sigma}})$.  In particular,  fiducial confidence limits for
$h(\bm\Sigma)$ will be constructed using corresponding percentiles of $h(\tilde{\bm{\Sigma}})$.

\renewcommand\thesubsection{S.\arabic{subsection}}
\section{Supplementary Material}

\subsection{Algorithm for Obtaining  $100(1-\alpha)\%$ Fiducial CI of $CCC_L$}
\begin{enumerate}
    \item For a given longitudinal data, fit the generalized linear model with a suitable link function in (1) in the manuscript using the integrated maximum likelihood method. Note down the estimates of fixed effects and variance components from the fitted model. These are consistent estimates by the properties of the integrated MLE.
    \item Calculate $\bm{y}^*_{il}$ using $\bm{y}_{il}$ in the simulated data and the estimates found from the model in Step 1. Calculate the marginal variance of $\bm{e}^*_{il}$ by plugging in the consistent estimates of variance components. Only for Steps 3, 4, and 5, treat this variance as known error variance in the LMM in  (8) in the manuscript.
    \item Compute the pseudo observations for this fiducial approach, $\hat{\bm{\mu}}_{\alpha_s i}$ and $\hat{\bm{\mu}}_{\gamma i}$ by plugging in the consistent estimates of variance component and are assumed to follow asymptotic normal distributions described in (9) in the manuscript.
    \item Using the method described in Section 4.2 in the manuscript, generate random variables from standard normal and chi-squares distributions to generate samples for fiducial pivot statistics of covariance matrix of the asymptotically jointly normally distributed $(\hat{\bm{\mu}}_{\alpha_0 i}, \hat{\bm{\mu}}_{\alpha_1 i}, ..., \hat{\bm{\mu}}_{\alpha_S i}, \hat{\bm{\mu}}_{\gamma i})$, which consists of the following components, $\bm{\sigma}^s_\alpha \bm{\Sigma}_{Y^*}^{-1} {\bm{\sigma}^{s'}_\alpha}^t$, $\bm{\sigma}^s_\alpha \bm{\Sigma}_{Y^*}^{-1} {\bm{\sigma}_\gamma}^t$ for $s, s'=0,...,S$, and $\bm{\sigma}_\gamma \bm{\Sigma}_{Y^*}^{-1} {\bm{\sigma}_\gamma}^t$. The analytical expressions for components of these matrices are functions of $\sigma_{\alpha ll\prime}^s$ for all $s=0,...,S$ and $\sigma_{\gamma ll\prime}$, the variance components of the GLMM (1) in the manuscript.
    \item For each generated sample for pivot statistics of $\bm{\sigma}^s_\alpha \bm{\Sigma}_{Y^*}^{-1} {\bm{\sigma}^{s'}_\alpha}^t$, $\bm{\sigma}^s_\alpha \bm{\Sigma}_{Y^*}^{-1} {\bm{\sigma}_\gamma}^t$ for $s, s'=0,...,S$ and $\bm{\sigma}_\gamma \bm{\Sigma}_{Y^*}^{-1} {\bm{\sigma}_\gamma}^t$, minimize the squared sum of the differences of analytical expression and the generated samples from their pivots and solve for $\sigma_{\alpha ll\prime}^s$ for all $s=0,...,S$ and $\sigma_{\gamma ll\prime}$ to obtain samples from their pivot statistics. The positive definiteness constraint on the covariance matrix needs to be handled using log-Cholesky decomposition. The minimization problem is addressed using the Inexact Newton Method (\cite{optimizationbook}) (see Section 4.3.1  in the manuscript for details).
    \item Generate $\bm{Z}_\beta$ from a standard normal distribution to get samples for pivot statistics of fixed effects using (10) in the manuscript after plugging in the pivot statistics for variance components in $\bm{\Sigma}_C$.
    \item If there is a closed-form expression of $CCC_L$, generate samples for pivot statistics of $CCC_L$ by plugging in all pivot statistics of fixed effects and variance components in the expression of $CCC_L$ found in Steps 4 and 5. Note that $CCC_L$ does not have any closed-form expression in general, but it can be computed numerically by involving numerical integration or Monte Carlo integration, given the values of the model parameters. The pivot statistics of $CCC_L$ can similarly be computed by plugging in the pivot statistics of the model parameters found in steps 4, 5 and 6 in the numerical method of computing $CCC_L$. Details have been discussed in Section 4.3 in the manuscript.
    \item Use a total of $10000$ samples generated using Steps 4, 5, 6, 7 to estimate the density curve of pivot distribution of $CCC_L$. Find the highest density region which covers $100(1-\alpha)\%$ of the total area under the simulated density curve based on the algorithm by \cite{kruschke2014}.
\end{enumerate}

\subsection{Algorithm for Obtaining Simulated Coverage Probability}
\begin{enumerate}
    \item Simulate $10,000$ datasets from the mentioned model, maintaining the sample size.
    \item Find fiducial confidence interval of $CCC_L$ for each  dataset.
    \item Calculate $c$, the number of confidence intervals that include the true value of $CCC_L$ among those $10,000$ confidence intervals. $\frac{c}{10,000}$ is computed as the simulated coverage probability.
\end{enumerate}

\subsection{A Proxy Approach for Finding Pivot Statistics of the Covariance Matrices of  Random Effects}
Note that $\hat{\bm{\mu}}_{\alpha_s i}$ for $i=1,...,N$ can be treated as $N$ i.i.d. observations from $L$-variate normal distributions in (9) in the manuscript for any $s=0,...,S$ and similarly $\hat{\bm{\mu}}_{\gamma i}$ for $i=1,...,N$ can be treated as $N$ i.i.d. observations from $L$-variate normal distribution in (10) in the manuscript.
Using the method described in Section 4.1 in the manuscript, component-wise pivot statistics can be obtained separately for each component of $\bm{\sigma}^s_\alpha \bm{\Sigma}_{Y^*}^{-1} {\bm{\sigma}^s_\alpha}^t$ for $s=0,...,S$ and $\bm{\sigma}_\gamma \bm{\Sigma}_{Y^*}^{-1} {\bm{\sigma}_\gamma}^t$. The analytical expressions of these covariance matrices are functions of variance components $\sigma_{\alpha ll\prime}^s$ for $s=0,...,S$ and $\sigma_{\gamma ll\prime}$ of $\bm{\Sigma}_\alpha^s$ for $s=0,...S$ and $\bm{\Sigma}_\gamma$ respectively. We equate these analytical expressions of variance and covariance components with their fiducial pivot statistics. This creates a system of non-linear equations; a total of  $L(L+1)(S+2)/2$ equations, ($L(L+1)/2$ from $\bm{\sigma}_\gamma \bm{\Sigma}_{Y^*}^{-1} {\bm{\sigma}_\gamma}^t$ and $(S+1)L(L+1)/2$ from $\bm{\sigma}^s_\alpha \bm{\Sigma}_{Y^*}^{-1} {\bm{\sigma}^s_\alpha}^t$ for $s=0,...,S$) and the same number of parameters for solution (a total $(S+1)L(L+1)/2$ $\sigma_{\alpha ll\prime}^s$ for $s=0,...,S$ and additional $L(L+1)/2$ $\sigma_{\gamma ll\prime}$). For solving this system of non-linear equations, the \textit{Broyden} algorithm has been used based on \cite{nleqslv}. Again, note that, the expressions can be function of dispersion parameter $\phi$. In such cases, $\phi$ must be replaced by its pivot statistic in the system of non-linear equations prior to solving those.

\noindent Let's consider a two-level design where $30$ subjects are observed longitudinally over $10$ time-points by two raters. We consider an LMM version of  (1) (expression is given in the manuscript) with a subject-specific random intercept ($\alpha_{0i}^{(l)}$, $l=1,2$) and random slope ($\alpha_{1i}^{(l)}$, $l=1,2$) for each rater and  $cov(\begin{pmatrix}
    \alpha_{0i}^{(1)}\\
    \alpha_{0i}^{(2)}
\end{pmatrix}) = \begin{pmatrix}
    0.45 & 0.40\\
    0.40 & 0.49
\end{pmatrix}$ and $cov(\begin{pmatrix}
    \alpha_{1i}^{(1)}\\
    \alpha_{1i}^{(2)}
\end{pmatrix}) = \begin{pmatrix}
    0.30 & 0.21\\
    0.21 & 0.18
\end{pmatrix}$. According to our assumptions for  (1), $\alpha_{0i}^{(l)}$ and $\alpha_{1i}^{(l)}$ are independent, $l=1, 2$. After obtaining the maximum likelihood estimates of these parameters, we obtain $\hat{\mu}_{si}^{(l)}$ as the estimate of the conditional mean of $\alpha_{si}^{(l)}$ given the observed data $\bm{y}$, $s=0, 1$, $l=1, 2$. Following the general notation, we denote $\hat{\bm{\mu}}_{\alpha_si} = \begin{pmatrix}
    \hat{\mu}_{si}^{(1)}\\
    \hat{\mu}_{si}^{(2)}
\end{pmatrix}$ for $s=0, 1$ ($0$ for random intercept and $1$ for random slope). We observe that each element of $cov(\hat{\bm{\mu}}_{\alpha_0i}, \hat{\bm{\mu}}_{\alpha_1i}) = \begin{pmatrix}
    cov(\hat{\mu}_{0i}^{(1)}, \hat{\mu}_{1i}^{(1)}) & cov(\hat{\mu}_{0i}^{(1)}, \hat{\mu}_{1i}^{(2)})\\
    cov(\hat{\mu}_{0i}^{(2)}, \hat{\mu}_{1i}^{(1)})
    & cov(\hat{\mu}_{0i}^{(2)}, \hat{\mu}_{1i}^{(2)})
\end{pmatrix} = \begin{pmatrix}
    0.00461 & 0.00093\\
    0.00094 & 0.00466
\end{pmatrix}$ is very small compared to the $var(\hat{\bm{\mu}}_{\alpha_0i})$ and $var(\hat{\bm{\mu}}_{\alpha_1i}))$.  As a result, the covariance matrix of the joint normal distribution of $(\hat{\bm{\mu}}_{\alpha_0i}, \hat{\bm{\mu}}_{\alpha_1i})$ is approximately a block diagonal matrix. We computed the fiducial confidence interval of CCC in two ways- (i) using the method that considered  the joint normal distribution of $(\hat{\bm{\mu}}_{\alpha_0i}, \hat{\bm{\mu}}_{\alpha_1i})$ which is our original method, and (ii) using the marginal normal distribution of $\hat{\bm{\mu}}_{\alpha_0i}$ and $\hat{\bm{\mu}}_{\alpha_1i}$ separately, which is the proxy to the original method. The upper and lower limits of the confidence interval differed by $0.001$ on an average, which is negligible.

\subsection{Numerical  Results for Comparing the Proposed Fiducial Method with that in \cite{cisewski2012} under the \cite{tsai2018} Model}
We have mentioned in the manuscript that the fiducial method for an LMM developed by \cite{cisewski2012} is for the univariate case, whereas we are dealing with a multivariate mixed effects model.  Thus it appears that the \cite{cisewski2012} fiducial method cannot be directly used for the model in equation (1) in the manuscript. We still want to compare our proposed method with that of \cite{cisewski2012}. For this we have used a model considered by \cite{tsai2018}, where we can apply both methods for obtaining  fiducial intervals for the CCC. We have used a simplified version of the model proposed by \cite{tsai2018}, given by
\begin{equation}
    y_{ijk} = \beta_0 + \beta_{rater}\text{Rater} + a_i + e_{ijk},
\end{equation}
where, $y_{ijk}$ denotes the measurement for subject $i$ at time-point $j$ by rater $k$, $i=1,...,n$, $j=1,...,T$, $k=1,2$.  Here we have considered a two rater scenario. Thus the covariate `Rater' is binary, taking 1 for rater 1 and 0 for rater 2. The random effect $a_i \widesim[1.5]{i.i.d.} N(0, \sigma^2_a)$ and the random error $e_{ijk} \widesim[1.5]{i.i.d.} N(0, \sigma^2_e)$. The CCC can now be expressed as $\frac{\sigma^2_a}{\sigma^2_a + \sigma^2_e/T + 0.5\beta_{rater}^2}$. In the following table, we have compared the fiducial 95\% confidence intervals for all model parameters as well as for the CCC in terms of average width and coverage probability. In the simulation setup, we have varied the sample size ($n$), but fixed the number of time points as $T=3$.
\begin{table}[h]
\renewcommand{\arraystretch}{1.5}
\fontsize{10}{10}\selectfont
\centering
\caption{Coverage probabilities and expected widths of 95\% CI by the proposed method and the method due to \cite{cisewski2012}}
\begin{tabular}{|c|c|c|c|c|c|c|}
\hline
Sample & Parameter & True & \multicolumn{2}{|c|}{Average 95\% CI} & \multicolumn{2}{|c|}{Coverage Probability}\\
\cline{4-7}
Size & & Value & Proposed Method & Cisewski \& Hannig & Proposed Method & Cisewski \& Hannig\\
($n$) & & & [Average Width] & [Average Width] & & \\
\hline
\multirow{5}{*}{5} & $\beta_0$ & 0.75 & (-0.45, 1.96) [2.41] & (-0.47, 1.98) [2.45] & 0.960 & 0.965 \\
 & $\beta_{rater}$ & 0.2 & (-0.05, 0.45) [0.50] & (-0.08, 0.47) [0.55] & 0.951 & 0.954 \\
 & $\sigma^2_a$ & 1 & (0.22, 5.91) [5.69] & (0.37, 8.90) [8.53] & 0.945 & 0.963 \\
 & $\sigma^2_e$ & 0.2 & (0.13, 0.30) [0.17] & (0.13, 0.33) [0.20] & 0.945 & 0.942 \\
 & CCC & 0.92 & (0.64, 0.97) [0.33] & (0.58, 0.99) [0.41] & 0.942 & 0.964 \\ \hline
\multirow{5}{*}{10} & $\beta_0$ & 0.75 & (0.05, 1.44) [1.39] & (0.04, 1.45) [1.41] & 0.947 & 0.948 \\
 & $\beta_{rater}$ & 0.2 & (0.03, 0.38) [0.35] & (0.02, 0.39) [0.37] & 0.948 & 0.952 \\
 & $\sigma^2_a$ & 1 & (0.36, 2.68) [2.32] & (0.46, 3.35) [2.89] & 0.946 & 0.957 \\
 & $\sigma^2_e$ & 0.2 & (0.15, 0.27) [0.12] & (0.15, 0.28) [0.13] & 0.947 & 0.947 \\
 & CCC & 0.92 & (0.79, 0.97) [0.18] & (0.76, 0.98) [0.22] & 0.946 & 0.957 \\ \hline
\multirow{5}{*}{30} & $\beta_0$ & 0.75 & (0.38, 1.14) [0.76] & (0.38, 1.15) [0.77] & 0.949 & 0.950 \\
 & $\beta_{rater}$ & 0.2 & (0.07, 0.33) [0.26] & (0.06, 0.33) [0.27] & 0.949 & 0.951 \\
 & $\sigma^2_a$ & 1 & (0.59, 1.67) [1.08] & (0.63, 1.85) [1.22] & 0.948 & 0.953 \\
 & $\sigma^2_e$ & 0.2 & (0.16, 0.25) [0.09] & (0.16, 0.26) [0.10] & 0.948 & 0.949 \\
 & CCC & 0.92 & (0.86, 0.96) [0.10] & (0.86, 0.97) [0.11] & 0.949 & 0.952 \\ \hline
\end{tabular}
\label{tab:comparison}
\end{table}
For the purpose of comparison, we are including results on the interval estimation of various parameters (in addition to the CCC) under the  model being considered; the results appear  in Table \ref{tab:comparison}. The numerical results show that overall, the \cite{cisewski2012} solution is somewhat conservative, resulting in wider confidence intervals. This is especially so for the subject-level random effect. For small sample sizes, our approach gives confidence intervals whose coverages are slightly lower than the assumed nominal level. For sample size 30 in Table \ref{tab:comparison}, both  approaches have similar coverages and similar expected widths for the confidence intervals of the various parameters. An exception is once again the subject-level random effect where our approach seems to have an edge in terms of the expected width. Focusing on the CCC, we note that for somewhat large sample sizes both methods give very similar results.

\subsection{Additional Simulated Results for a Mixture and Gaussian Models}

\begin{table}[h]
    \caption{\textbf{Robust performance  of the  fiducial approach in comparison with Fisher Z and bootstrap}} 
    \centering
    \vspace{.4cm}
\fontsize{8.5}{10}\selectfont
    \centering
     \begin{tabular}{|c|c|c|c|c|c|c|}
    \hline
      & CCC & Sample & Method & Average 95\% & Expected Width & Inclusion\\
    Distribution of $\epsilon$ & (True) & Size & & Confidence & of 95\% & Probability\\
    &        & (N) & & Limits & Confidence Limits & \\
     \hline
     \multirow{9}{*}{
     \fontsize{7.5}{10}\selectfont
  \begin{tabular}{c}
     $\epsilon \sim 0.90 Normal + 0.10 Gamma$\\
     $Normal$ with mean 0 and variance 0.11\\
     $Gamma$ (centered) with\\ 
     shape 0.5 and scale 2\\
     Skewness of $Gamma$ = 2.83\\
     Excess Kurtosis of $Gamma$ = 12
  \end{tabular}} & \multirow{9}{*}{0.801} & \multirow{3}{*}{30} & Fiducial & (0.681, 0.880) & \textbf{0.199} & \textbf{0.924}\\
    & & & Fisher Z transform & (0.621, 0.887) & 0.266 & 0.981\\
    & & & Bootstrap & (0.651, 0.890) & 0.239 & 0.910\\
    \cline{3-7}
    & & \multirow{3}{*}{50} & Fiducial & (0.711, 0.863) & \textbf{0.152} & \textbf{0.934}\\
    & & & Fisher Z transform & (0.678, 0.874) & 0.196 & 0.985\\
    & & & Bootstrap & (0.690, 0.871) & 0.181 & 0.918\\
    \cline{3-7}
    & &  \multirow{3}{*}{100} & Fiducial & (0.742, 0.846) & \textbf{0.106} & \textbf{0.941}\\
    & & & Fisher Z transform & (0.715, 0.852) & 0.137 & 0.984\\
    & & & Bootstrap & (0.729, 0.851) & 0.122 & 0.929\\
     \hline
     \multirow{9}{*}{
  \fontsize{7.5}{10}\selectfont
  \begin{tabular}{c}
    $\epsilon \sim 0.90 Normal + 0.10 LogNormal$\\
     $Normal$ with mean 0 and variance 0.11\\
     $LogNormal$ (centered) with\\ 
     location 0.5 and scale 0.7\\
     Skewness of $Z$ = 2.88\\
     Excess Kurtosis of $Z$ = 18
  \end{tabular}} & \multirow{9}{*}{0.800} & \multirow{3}{*}{30} & Fiducial & (0.661, 0.872) & \textbf{0.211} & \textbf{0.927}\\
    & & & Fisher Z transform & (0.612, 0.894) & 0.282 & 0.979\\
    & & & Bootstrap & (0.646, 0.887) & 0.241 & 0.908\\
    \cline{3-7}
    & & \multirow{3}{*}{50} & Fiducial & (0.709, 0.862) & \textbf{0.153} & \textbf{0.930}\\
    & & & Fisher Z transform & (0.662, 0.877) & 0.215 & 0.982\\
    & & & Bootstrap & (0.687, 0.873) & 0.186 & .915\\
    \cline{3-7}
    & & \multirow{3}{*}{100} & Fiducial & (0.737, 0.844) & \textbf{0.107} & \textbf{0.935}\\
    & & & Fisher Z transform & (0.707, 0.858) & 0.151 & 0.980\\
    & & & Bootstrap & (0.731, 0.856) & 0.125 & 0.920\\
     \hline
    \end{tabular}
\end{table}
For Table S.1, a two level design with two raters has been used (similar to the design in Table 1). A mixture  distribution with errors 90\% from a normal (mean=0)   and 10\%  from  a  heavy tailed skewed distribution was used for the simulation. Parameters are set to : $\beta_0^{(1)} = 0.75$, $\beta_0^{(2)} = 0.50$, $\beta_1^{(1)} = -0.10$, $\beta_1^{(2)} = -0.06$, $\begin{pmatrix} \sigma^0_{\alpha11} & \sigma^0_{\alpha12}\\
    \sigma^0_{\alpha12} & \sigma^0_{\alpha22} \end{pmatrix} = 
    \begin{pmatrix} 0.45 & 0.40\\ 0.40 & 0.49 \end{pmatrix}$, 
    $\begin{pmatrix} \sigma^1_{\alpha11} & \sigma^1_{\alpha12}\\
    \sigma^1_{\alpha12} & \sigma^1_{\alpha22} \end{pmatrix} = 
    \begin{pmatrix} 0.30 & 0.20\\ 0.20 & 0.18 \end{pmatrix}$. Each subject is observed over 10 time-points, i.e. $T=10$.\\
    
\noindent Inspecting Table S.1, we find that for all parametric combinations, fiducial under performs slightly for n=30, however its performance gradually improves over n. Fisher Z is still liberal in terms of the coverage probability and as a result, its expected width is larger than that of the fiducial. Underperformance of bootstrap is revealed everywhere in  Table S.1  in terms of the expected width and coverage probability.\\

\begin{table}[h]
\caption{\textbf{Fiducial performance of a two-level design with three raters}} 
    \centering 
    \vspace{.4cm}
    \renewcommand{\arraystretch}{1.5}
\fontsize{10}{10}\selectfont
    \begin{tabular}{|c|c|c|c|c|}
    \hline
    CCC (True) & Sample Size (N) & Average 95\% & Expected Width of & Inclusion Probability\\
    & & Confidence Limit & 95\% Confidence Limit &\\
    \hline
    \multirow{3}{*}{0.731} & 30 & (0.562, 0.820) & 0.258 & 0.930\\
    & 50 & (0.601, 0.800) & 0.199 & 0.943\\
    & 100 & (0.647, 0.785) & 0.138 & 0.952\\
    \hline
    \end{tabular}
\end{table}

\noindent For Table S.2, a two-level design with three raters has been used for this simulation. Under this setting, the expression of CCC in (15) becomes:\\
$\frac{2(\sigma_{\alpha 12}^0 + \sigma_{\alpha 13}^0 + \sigma_{\alpha 23}^0 + \sum \limits_{j=1}^T (\sigma_{\alpha 12}^1 + \sigma_{\alpha 13}^1 + \sigma_{\alpha 23}^1)j^2}{\sigma_{\alpha 11}^0 + \sigma_{\alpha 22}^0 + \sigma_{\alpha 33}^0 + \sum \limits_{j=1}^T (\sigma_{\alpha 11}^1 + \sigma_{\alpha 22}^1 + \sigma_{\alpha 33}^1)j^2 + 3 \sigma^2 + \frac{1}{T} \sum \limits_{j=1}^T \sum \limits_{l=1}^3 \sum \limits_{l^\prime = l+1}^3 \left[(\beta^{(l)}_0 - \beta^{(l^\prime)}_0) + (\beta^{(l)}_1 - \beta^{(l^\prime)}_1)j\right]^2}$. The parameters used for simulations are: $\beta_0^{(1)} = 0.75$, $\beta_0^{(2)} = 0.50$, $\beta_0^{(3)} = 0.60$, $\beta_1^{(1)} = -0.10$, $\beta_1^{(2)} = -0.06$, $\beta_1^{(2)} = -0.08$, \\
$\begin{pmatrix} \sigma^0_{\alpha11} & \sigma^0_{\alpha12} & \sigma^0_{\alpha13}\\
\sigma^0_{\alpha12} & \sigma^0_{\alpha22} & \sigma^0_{\alpha23}\\
\sigma^0_{\alpha13} & \sigma^0_{\alpha23} & \sigma^0_{\alpha33}
\end{pmatrix} = \begin{pmatrix} 0.45 & 0.40 & 0.42\\ 0.40 & 0.49 & 0.40\\
0.42 & 0.40 & 0.45 \end{pmatrix}$, $\begin{pmatrix} \sigma^1_{\alpha11} & \sigma^1_{\alpha12} & \sigma^1_{\alpha13}\\
\sigma^1_{\alpha12} & \sigma^1_{\alpha22} & \sigma^1_{\alpha23}\\
\sigma^1_{\alpha13} & \sigma^1_{\alpha23} & \sigma^1_{\alpha33}
\end{pmatrix} = \begin{pmatrix} 0.10 & 0.067 & 0.05\\ 0.067 & 0.06 & 0.06\\
0.05 & 0.06 & 0.08 \end{pmatrix}$.\\

Each subject is observed over 10 time points, i.e. $T=10$.
\noindent Inspecting Table S.2, we find that for the specified parametric combination, fiducial performs extremely well for all n=30, 50, and 100. Expected widths get tighter and inclusion probabilities become larger as n increases. Particularly, for n=100, the inclusion probability is more than 0.95 proving that our approach does not have a  downward bias.

\begin{table}[h]
\caption{\textbf{Fiducial performance of a three level design with two raters}} 
    \centering 
    \vspace{.4cm}
    \renewcommand{\arraystretch}{1.5}
\fontsize{10}{10}\selectfont
    \begin{tabular}{|c|c|c|c|c|}
    \hline
    CCC (True) & Sample Size (N) & Average 95\% & Expected Width of & Inclusion Probability\\
    & & Confidence Limit & 95\% Confidence Limit &\\
    \hline
    \multirow{3}{*}{0.772} & 15 & (0.613, 0.874) & 0.261 & 0.939\\
    & 30 & (0.669, 0.847) & 0.178 & 0.948\\
    & 50 & (0.695, 0.832) & 0.137 & 0.951\\
    \hline
    \end{tabular}
\end{table}

\noindent For Table S.3, a three level design with two raters has been used for  simulation (excluding the subject-level random slope in (14)). Under this setting, the expression of CCC   becomes:\\
$\frac{2(\sigma_{\alpha 12}^0 + \sigma_{\gamma 12})}{\sigma_{\alpha 11}^0 + \sigma_{\alpha 22}^0 + \sigma_{\gamma 11} + \sigma_{\gamma 22} + 2 \sigma^2 + \frac{1}{T} \sum \limits_{j=1}^T \left[(\beta^{(1)}_0 - \beta^{(2)}_0) + (\beta^{(1)}_1 - \beta^{(2)}_1)j\right]^2}$. Parameters used for simulations are: $\beta_0^{(1)} = 0.65$, $\beta_0^{(2)} = 0.40$, $\beta_1^{(1)} = -0.10$, $\beta_1^{(2)} = -0.06$,
$\begin{pmatrix} \sigma^0_{\alpha11} & \sigma^0_{\alpha12}
\sigma^0_{\alpha12} & \sigma^0_{\alpha22}\end{pmatrix} =
\begin{pmatrix} 0.45 & 0.44\\ 0.44 & 0.49 \end{pmatrix}$, $\begin{pmatrix} \sigma^0_{\gamma11} & \sigma^0_{\gamma12}\\
\sigma^0_{\gamma12} & \sigma^0_{\gamma22}
\end{pmatrix} = \begin{pmatrix} 0.10 & 0.067\\ 0.067 & 0.06 \end{pmatrix}$. Each subject is observed over 10 time points, i.e. $T=10$ and at each time point, 5 observations are taken for each subject, i.e. $K=5$.  Table S.3 basically mimics Table S.2 with a slight better result. Thus, we feel very comfortable to use the fiducial approach for both two and three level designs.

\subsection{Simulation Results for a Gamma Family}
\noindent  For right skewed ratings, we use the GLMM setup under the assumption that  errors follow a gamma distribution. For the gamma GLMM, several link functions are being used in the literature, such as log link, square root link, inverse link, etc. We consider the inverse links for this discussion, mainly motivated by the complexity of computing the expression of CCC. $CCC_L$ is computed using the expression in (4) in the manuscript, where integrations involved in computing expectation, variance, and covariance terms are approximated numerically by Monte Carlo integration.

\noindent Based on the gamma family, the GLMM in (1) in the manuscript can be modified by taking $h(\eta) = \eta^{-1}$. This implies $\bm{\mu} = \bm{\eta}^{-1}$, thus $\mathcal{G}^{\prime}(\bm{\mu}) = [diag(-\bm{\eta}^{-2})]^{-1} = diag(-\bm{\mu}^{-2})$ and $\bm{y}^* = \hat{\bm{\eta}} - diag(\hat{\bm{\mu}}^{-2})(\bm{y} - \hat{\bm{\mu}})$ in (8) in the manuscript and additionally, the dispersion parameter, $\phi = 1$ and $\zeta(\mu) = \frac{\mu^2}{\tau}$, where $\tau$ is the shape parameter of the gamma distribution. Based on (7) in the manuscript, conditioned on random effects, $var(\bm{\epsilon}) = \tau^{-1}diag(\bm{\mu}^2)$, this implies, $\bm{\epsilon}^* \sim N(\bm{0}, \tau^{-1}diag(\bm{\mu}^{-2}))$ in (8) in the manuscript. For the simulation purpose, instead of generalizing the model in (8) in the manuscript, we have used a more specific version of (8),   same as Gaussian and Poisson examples, which is described below for a rater $l$,

\begin{equation*}
    E(y_{ijkl}|\alpha^{(l)}_{0i}, \alpha^{(l)}_{1i}, \gamma^{(l)}_{0ij})= \mu_{ijkl} = \left(\beta^{(l)}_{0}+ \beta^{(l)}_{1}j + \alpha^{(l)}_{0i} + \alpha^{(l)}_{1i}j + \gamma^{(l)}_{0ij}\right)^{-1}, \quad l=1,...,L.
\end{equation*}
After transforming $y_{ijkl}$ to $y_{ijkl}^*$ as $y_{ijkl}^* = \frac{1}{\hat{\mu}_{ijkl}} - \frac{1}{\hat{\mu}_{ijkl}^2}(y_{ijkl} - \hat{\mu}_{ijkl})$, the transformed model in (8) in the manuscript can be simplified for this example as,
\begin{equation*}
    y_{ijkl}^*= \beta^{(l)}_{0}+ \beta^{(l)}_{1}j + \alpha^{(l)}_{0i} + \gamma^{(l)}_{0ij} + \epsilon_{ijkl}^*, \quad \quad l=1,...,L,
\end{equation*}
which can be jointly written for all raters as,
$$y_{ijkl}^* = \sum \limits_{m=1}^L \beta^{(l)}_{0}\delta_{(m=l)} + \sum \limits_{m=1}^L\beta^{(l)}_{1}\delta_{(m=l)}j + \sum \limits_{m=1}^L\alpha^{(l)}_{0i}\delta_{(m=l)} + \sum \limits_{m=1}^L\gamma^{(l)}_{0ij}\delta_{(m=l)} + \epsilon_{ijkl}^*,
$$
where, $\epsilon_{ijkl}^*|\alpha^{(l)}_{0i}, \alpha^{(l)}_{1i}, \gamma^{(l)}_{0ij} \widesim[1.5]{indp} N(0, \frac{1}{\tau\mu_{ijkl}^2})$ for $i=1,...,N$, $j=1,...,T$, $k=1,...,K$, $l=1,...,L$. The details of the model are similar to the Gaussian and Poisson examples discussed in the previous subsections.\\
\noindent A relatively large value of $\tau=25$  has been considered to achieve a meaningful value of $CCC_L$ for this simulation, which also resembles the real data example in Section 5. Theoretically, a larger value of $\tau$ makes the gamma family close to a normal family, but we have verified that a gamma family with $\tau=25$ significantly differs from a normal family with a p-value less than $10^{-16}$ using the Shapiro-Wilk normality test (\cite{royston1982}).\\
As the closed-form expression of CCC is very difficult to obtain for this model, the Numerical CCC approach (explained in Section 4.2 in the manuscript) has been used here to compute the fiducial confidence intervals. From the results in the Poisson example (Table 2, Section 4.2 in the manuscript), it is convincing enough that the Numerical CCC approach produces as good results as the Exact CCC approach. No wonder, this has also been reflected in Table S.4, as we observe that the fiducial confidence intervals almost achieved the target inclusion probability even for a sample, as small as 30.

\begin{table}[h]
\caption{\textbf{Fiducial performance for a three level design with a gamma distribution}} 
    \centering 
    \vspace{.4cm}
    \renewcommand{\arraystretch}{1.5}
\fontsize{10}{10}\selectfont
    \begin{tabular}{|c|c|c|c|c|}
    \hline
    CCC (True) & Sample Size (N) & Average 95\% & Expected Width of & Inclusion Probability\\
    & & Confidence Limit & 95\% Confidence Limit &\\
    \hline
    \multirow{4}{*}{0.622} & 15 & (0.453, 0.747) & 0.294 & 0.928\\
    & 30 & (0.501, 0.734) & 0.233 & 0.957\\
    & 50 & (0.536, 0.725) & 0.189 & 0.953\\
    & 100 & (0.559, 0.701) & 0.142 & 0.949\\
    \hline
    \end{tabular}
\end{table}
\noindent For Table S.4, a three-level design with two raters has been considered for the simulation. Using the given model, the expression of $CCC_L$ has been computed by the  Monte Carlo integration using fiducial statistics for model parameters.
Parameters used for simulations are : $\beta_0^{(1)} = 2$, $\beta_0^{(2)} = 1.90$, $\beta_1^{(1)} = 0.08$, $\beta_1^{(2)} = 0.06$, and \\

$\begin{pmatrix} \sigma^0_{\alpha11} & \sigma^0_{\alpha12}\\
\sigma^0_{\alpha12} & \sigma^0_{\alpha22}
\end{pmatrix} = \begin{pmatrix} 0.25 & 0.22\\ 0.22 & 0.20 \end{pmatrix}$, $\begin{pmatrix} \sigma^0_{\gamma11} & \sigma^0_{\gamma12}\\ \sigma^0_{\gamma12} & \sigma^0_{\gamma22} \end{pmatrix} = \begin{pmatrix} 0.05 & 0.04\\ 0.04 & 0.04 \end{pmatrix}$, $\tau = 25$. Each subject is observed over 10 time points, i.e. $T=10$, and at each time point, 5 observations are taken for each subject by each of 2 raters, i.e., $K=5$, $L=2$. Inspecting Table S.4, we see that the target coverage probability of $95\%$ is attained for all $n=30, 50, 100$ and the  width of the confidence interval gets narrower as $n$ increases.

\bibliographystyle{agsm2}
\bibliography{references}